\documentclass[useAMS,usenatbib]{mn2e}
\usepackage{graphics}
\usepackage{subfigure}
\usepackage{graphicx}
\usepackage{amssymb}
\usepackage{hyperref}

\def \sag{{\small SAG}}
\def \fof {{\small FOF}}

\begin{document}

% If your system does not have the AMS fonts version 2.0 installed, then
% remove the useAMS option.
%
% useAMS allows you to obtain upright Greek characters.
% e.g. \umu, \upi etc.  See the section on "Upright Greek characters" in
% this guide for further information.
%
% If you are using AMS 2.0 fonts, bold math letters/symbols are available
% at a larger range of sizes for NFSS release 1 and 2 (using \boldmath or
% preferably \bmath).
%
% The usenatbib command allows the use of Patrick Daly's natbib.sty for
% cross-referencing.
%
% If you wish to typeset the paper in Times font (if you do not have the
% PostScript Type 1 Computer Modern fonts you will need to do this to get
% smoother fonts in a PDF file) then uncomment the next line
% \usepackage{Times}

%%%%% AUTHORS - PLACE YOUR OWN MACROS HERE %%%%%
   
\title[The bright end of the CMR]{The bright end of the colour-magnitude relation of cluster
  galaxies} \author[Noelia Jim\'enez et al.]{Noelia
  Jim\'enez$^{1,2}$\thanks{E-mail: njimenez@fcaglp.unlp.edu.ar},
  Sof\'ia A. Cora$^{1,2}$, Lilia P. Bassino$^{1,2}$, 
  Tom\'as E. Tecce$^{2,3}$ and
  \newauthor Anal\'ia V. Smith Castelli$^{1,2}$ \\
  $^{1}$Facultad de Ciencias Astron\'omicas y
  Geof\'isicas de la Universidad Nacional de La Plata, and Instituto
  de Astrof\'isica de La Plata\\ (CCT La Plata, CONICET, UNLP),
  Observatorio Astron\'omico, Paseo del Bosque S/N, B1900FWA La Plata,
  Argentina\\ $^{2}$Consejo Nacional de Investigaciones Cient\'ificas
  y T\'ecnicas, Rivadavia 1917, C1033AAJ Buenos Aires,
  Argentina\\ $^{3}$Instituto de Astronom\'ia y F\'isica del Espacio,
  CC. 67 Suc. 28, C1428ZAA Ciudad de Buenos Aires, Argentina}

\date{Accepted. Received; in original form}

\pagerange{\pageref{firstpage}--\pageref{lastpage}} \pubyear{2010}
\maketitle
\label{firstpage}
\begin{abstract}
  We investigate the physical processes involved in the development of
  the red sequence (RS) of cluster galaxies by using a combination of
  cosmological {\em N}-body simulations of clusters of galaxies and a
  semi-analytic model of galaxy formation. Results show good agreement
  between the general trend of the simulated RS and the observed
  colour-magnitude relation (CMR) of early-type galaxies in different
  magnitude planes. However, in many clusters, the most luminous
  galaxies ($M_R \sim M_V \sim M_{T_1} \lesssim -20$) depart from the
  linear fit to observed data, as traced by less luminous ones,
  displaying almost constant colours.  With the aim of understanding
  this particular behaviour of galaxies in the bright end of the RS,
  we analyze the dependence with redshift of the fraction of stellar
  mass contributed to each galaxy by different processes, i.e.,
  quiescent star formation, and starbursts triggered by disc
  instability and merger events. We find that the evolution of
  galaxies in the bright end since $z\approx 2$ is mainly driven by
  minor and major dry mergers, while minor and major wet mergers are
  relevant in determining the properties of less luminous galaxies.
  Since the most luminous galaxies have a narrow spread in ages
  ($1.0\times 10^{10}$ yr $<t<1.2\times 10^{10}$ yr), their
  metallicities are the main factor that affects their colours.  Their
  mean iron abundances are close to the solar value and have already
  been reached at $z \approx 1$. This fact is consistent with several
  observational evidences that favour a scenario in which both the
  slope and scatter of the CMR are in place since $z \approx 1.2$.
  Galaxies in the bright end reach an upper limit in metallicity as a
  result of the competition of the mass of stars and metals provided
  by the star formation occuring in the galaxies themselves and by the
  accretion of merging satellites.  Star formation activity in massive
  galaxies (stellar mass $M_\star \gtrsim 10^{10} M_{\odot}$) that
  takes place at low redshifts contribute with stellar components of
  high metallicity, but the fraction of stellar mass contributed since
  $z\approx 1$ is negligible with respect to the total mass of the
  galaxy at $z=0$.  On the other hand, mergers contribute with a
  larger fraction of stellar mass ($\approx 10-20$ per cent), but the
  metallicity of the accreted satellites is lower by $\approx 0.2$ dex
  than the mean metallicity of galaxies they merge with. The effect of
  dry mergers is to increase the mass of galaxies in the bright end,
  without significantly altering their metallicities. Hence, very
  luminous galaxies present similar colours that are bluer than those
  expected if recent star formation activity were higher, thus giving
  rise to a break in the RS.  These results are found for simulated
  clusters with different virial masses ($10^{14} -
  10^{15}\,h^{-1}\,{\rm M}_\odot$), supporting the idea of the
  universality of the CMR in agreement with observational results.
\end{abstract}


\section{Introduction}\label{Introsec}
It is now well established that galaxies follow a bimodal distribution
in the colour-magnitude plane, separated into a tight `red sequence'
(RS) and a `blue cloud'. The RS mostly consists of gas-poor galaxies
with low levels of star formation (SF), prototypically early-type
galaxies (ETGs), whereas late-type galaxies are typical objects of the
blue cloud.

The colour-magnitude relation (CMR) of ETGs can be understood as a
mass-metallicity relation. The more luminous (and therefore more
massive) galaxies in this relation have deep potential wells, capable
of retaining the metals released by supernovae events and stellar
winds. The CMR seems to be quite universal, since it is followed by
galaxies in the field as well as in groups and clusters
\citep[e.g.][]{Visvanathan77,Sandage78,Reda04,LopezCruz04,Reda05,deRijcke09},
but with a larger fraction of red galaxies being in denser
environments.

There is controversy on which are the mechanisms responsible for the
evolution of galaxies to the RS. The physical processes invoked can be
classified in internal and external. The former include passive
stellar evolution and disc instabilities, while the latter involve
galaxy-galaxy interactions and mergers \citep{Lin10,Robaina10}, and
environmental effects such as fast gravitational interactions (`galaxy
harassment', \citealt{Moore99}), removal of the hot gas reservoir
(`strangulation' or `starvation',
\citealt{Larson80,Balogh00,Kawata08}) and ram pressure stripping of
galactic gas \citep{Lanzoni05, Roediger09,Tecce10}.

Studying the evolution of the slope and scatter of the CMR provides
constraints on the SF history of ETGs. \citet{Bower98} find that the
observed CMR scatter in local clusters can be reproduced if galaxies
are assumed to form the bulk of their stellar content at early epochs
($z > 1$), with a subsequent moderate mass growth driven by merger
events and residual SF activity. This scenario is consistent with
hierarchical models of galaxy formation. In this context, using
semi-analytical modelling, \citet{Menci08} obtain a narrow RS for
cluster galaxies, already defined by $z \approx 1.2$. At a similar
redshift, \citet{Kaviraj05} find that the slope of the CMR of cluster
galaxies changes appreciably with respect to its present value,
although the evolution of the slope in the redshift range $0 < z <
0.8$ is negligible taking into account the errors.

Detailed observations of the CMR of cluster galaxies (see
\citealt{Mei09}, and references therein) show no significant evolution
of the zero point, slope and scatter of the CMR out to $z\approx
1.3$. In contrast with these results, \citet{Stott09} find from
optical and near-infrared observations of cluster galaxies an
evolution of the CMR slope between $z \approx 0.5$ and the present
epoch, which they attribute to the infall of galaxies into the cluster
core that are being transformed into RS galaxies.

Hydrodynamical simulations have also been used to analyse the
properties of the CMR at $z=0$. \citet{Saro06} evaluate the impact of
the stellar initial mass function (IMF), finding that the Salpeter IMF
allows to recover both the slope and the normalization of the CMR for
galaxies in cluster-sized haloes. From the analysis of simulations of
galaxy groups and clusters, \citet{Romeo08} conclude that the shape of
the RS is mainly determined by the specific SF in all
environments. These authors find that evolving galaxies move towards a
`dead' sequence soon after they have almost stopped the bulk of their
SF. Fainter galaxies in clusters keep a significant amount of SF out
to very recent epochs, and are thus distributed more broadly around the RS.
 
A linear relation has generally been used to fit the correlation
between luminosity and colour of the CMR, from giant to dwarf
ellipticals. However, the scatter of these relations increases at
lower luminosities, as becomes evident from observations of several
clusters (Coma, \citealt{Secker97}; Perseus, \citealt{Conselice03};
Fornax, \citealt{Hilker03}, \citealt{Karick03}, \citealt{Mieske07};
Hydra, \citealt{Misgeld08}; Virgo, \citealt{Lisker08}; Antlia,
\citealt{Analia08}).
 
Some of these relations seem to be consistent with a change of slope
from the bright to the faint end. As an example, this kind of fit has
already been suggested by \citet{Vacu61} for the Virgo cluster, and
confirmed later by \citet[e.g.][]{LaFerrarese06}. More recently, based
on Sloan Digital Sky Survey (SDSS) imaging data, \citet{Janz09} found,
also for Virgo, a non-linear relation that can be described by a `S'
shape over the whole range of magnitudes, with the brightest galaxies
($-21 \lesssim M_{\rm B} \lesssim -19)$ having an almost constant
colour. In the case of the Hydra cluster, \citet{Misgeld08} fit a
linear relation to its CMR, but a change of slope is evident for the
brightest galaxies in that cluster.

Additional evidence of a tilt towards bluer colours at the bright end
of the RS arises from studies of large samples of galaxies in the SDSS
\citep[e.g.][]{Baldry04,Baldry06,Skelton09}. The results of
\citet{Skelton09} were obtained for the local RS averaged over all
environments. \citet{Baldry06} examined the dependence of the fit
(using a `S'-shaped combination of straight line plus a hyperbolic
tangent function) to the mean positions of both the RS and blue cloud,
and found that unlike the fractions of red galaxies, the fits to the
RS do not vary strongly with environment.

The detachment of the bright end from the general trend denoted by the
linear fit to the CMR of cluster galaxies motivates our study. The CMR
constitutes one of the major tools to test galaxy formation models
since  galactic evolution is evidenced by it. Our aim is to contribute
to the understanding of the physical processes behind the development
of the RS in galaxy clusters, and the special behaviour of its bright
end.

Recent observations support strong evolution with cosmic time in the
structural properties of the most massive (stellar mass $M_\star
\gtrsim 10^{11} M_\odot$) spheroid-like galaxies. Size measurements of
ellipticals at high redshift ($z \gtrsim 1.5$) reveal that these
objects are much smaller (by factors $\sim 2$ to $\sim 4$) than their
local counterparts of comparable stellar mass
(\citealt{Daddi05}). This result was subsequently confirmed by several
studies, although the most massive ETGs ($M_\star \gtrsim
2.5\times10^{11} M_\odot$) seem to have reached their internal and
dynamical structure earlier and faster than lower mass ETGs at the
same redshift, as discussed by \citet{Mancini10}.  On the other hand,
a mild evolution in velocity dispersion is found for massive galaxies
\citep{CenarroTrujillo09}.  On average, the velocity dispersion and
surface mass density of massive galaxies at high redshift are similar
to those of the most dense local ETG \citep{Cappellari09}.  These
observational evidences suggest that inside-out growth scenarios are
plausible, in which the compact high redshift galaxies make up the
centres of normal nearby ellipticals \citep{vanDokkum10}.

From samples of galaxy pairs obtained from different surveys, it is
found that present-day massive ETGs undergo, on average, $\sim 0.5$
major mergers since $z \sim 0.6 - 0.7$
\citep{Bell06b,Robaina10}. These mergers generally occur between
galaxies which are already on the RS, thus involving ETGs containing
little cold gas and dust \citep{WhitakervanDokkum08,McIntosh08}. These
dissipationless (gas-poor, thus called `dry') major mergers are an
important channel for the formation and evolution of the brightest
cluster galaxies (BCGs) \citep{Liu09}. However, the rate of dry major
mergers is too low to fully explain the formation of spheroidal (and
RS) galaxies since $z\sim 1$ \citep{Bundy09}.  Minor dry mergers would
favour the growth of BCGs as inferred from the evolution of their
velocity dispersion-stellar mass relation \citep{Bernardi09}. The
minor merger hypothesis also appears to be plausible for early-type
galaxies at intermediate redshifts ($0.1<z<0.8$) as shown by
\citet{Nierenberg11} who find that ETGs are typically surrounded by
several satellite galaxies.  Minor mergers are also probably
responsible for the size evolution of massive ETGs
\citep{BezansonvanDokkum10}.  Besides, \citet{Kaviraj11} demonstrate
that the low-level of star formation activity in the ETG population at
intermediate redshift is likely to be driven by minor mergers.

The influence of mergers on the evolution of ETGs inferred from
observational data is in agreement with predictions obtained from
models of galaxy formation. Using semi-analytic modelling techniques,
\citet{KhochfarBurkert03} find that the last major merger of
present-day bright elliptical galaxies ($M_B \lesssim -21$) occurs
preferentially between bulge-dominated galaxies. Major mergers are
responsible for doubling the stellar mass at the massive end of the RS
since $z\sim 1$ \citep{Moral10}.  On the other hand,
\citet{Bournaud07} study the effect of multiple minor mergers on ETG
evolution by using a {\em N}-body simulation finding that repeated
minor mergers can form elliptical galaxies without major mergers
events, which are less frequent than minor mergers at moderate
redshift. Remnants of these multiple mergers develop global properties
that depend more on the total mass accreted during mergers than on the
mass ratio of each merger.  This result is also supported by
hydrodynamic simulations \citep[e.g.][]{Naab07,Naab09} which suggest
that the growth of massive galaxies ($M_\star \gtrsim 10^{11}
M_\odot$) by a large series of minor mergers becomes more important
than the growth via major mergers; minor mergers may thus be the main
driver for the late evolution of sizes and densities of ETGs. Another
possible physical mechanism responsible for the size evolution of ETGs
is the expulsion of a substantial fraction of the initial baryons,
still in gaseous form, by quasar activity \citep{Fan10}.

From semi-analytic modelling, \citet{Khochfar06b} find that massive
galaxies at high redshifts form in gas-rich mergers, whereas galaxies
of the same mass at low redshifts form from gas-poor mergers. These
considerations allow them to reproduce the observed size-redshift
evolution of massive spheroidal objects. Dissipation associated with
large gas fractions is the most important factor determining spheroid
sizes, as found by \citet{Hopkins09} using high resolution
hydrodynamic simulations. Complementarily, also using semi-analytic
models, \citet{DeLucia07} find that later ($z < 0.5$) minor dry
mergers play an important role in the mass assembly of BCGs, which
acquire half of their mass via accretion of smaller galaxies. 

It is clear that different types of mergers have a strong impact on
the evolution of ETGs, affecting their morphological and kinematical
properties, as well as their star formation history. As has been
described, many works have been devoted to study these aspects but
only a few have analysed in detail the role of mergers in the
development of the CMR \citep[e.g.][]{Skelton09, Bernardi10}. In
particular, by using a toy model combined with galaxy merger trees
extracted from a semi-analytic model, \citet{Skelton09} find that the
tilt towards bluer colours of the bright end of the RS  of local
early-type galaxies can be explained by the higher fraction of dry
mergers suffered by bright galaxies with respect to fainter
ones. Galaxy colour is not expected to change during dry mergers,
since there is no associated SF. Galaxies then move along the CMR as
the mass of the system increases, with their colour remaining fixed
\citep{Bernardi07}.

In this work, we investigate this issue in detail using the
semi-analytic model of galaxy formation and evolution
\sag~\citep*[acronym for Semi-Analytic Galaxies;][hereafter
  LCP08]{Lagos08}, which is combined with hydrodynamical cosmological
simulations of galaxy clusters by \citet{Dolag05}. We focus our study
on galaxies lying on the RS, which are selected by a colour
criterion. The samples thus obtained mostly include early type
galaxies, making our conclusions valid for the understanding of the
development of the CMR of ETGs. Thus, we hereafter use the term RS to
refer to colour-selected galaxies from our simulations, and the term
CMR to refer to observed relations for ETGs.

The tilt in the bright end of the RS emerges naturally from the model
once it has been calibrated to recover numerous observed galaxy
properties, both locally and at high redshift. We quantify the
contribution to the mass and metal content of galaxies located in the
RS given by quiescent SF from the cold gas available in the galaxy
disc, starbursts triggered by disc instability and merger events, and
the stellar mass accreted from satellite galaxies during mergers,
which are identified as minor/major and wet/dry ones.

This paper is organised as follows. In Section~\ref{Sec:model}, we
briefly describe the semi-analytic model used in this work and give
details of the cluster simulations. Section~\ref{Sec:RedSeq} presents
the comparison between the simulated RS and observed CMR followed by
early-type cluster galaxies in different magnitude planes, and the
analysis of the luminosity-metallicity relation followed by galaxies
in the RS. Section~\ref{Sec:Proc} describes the formalism used to
quantify the contribution of different processes to the mass and
metallicity of galaxies; the corresponding results are presented and
discussed.  Finally, in Section~\ref{Sec:Conclu} we summarize our main
findings.

\section{Hybrid Model of Galaxy Formation and Evolution}
\label{Sec:model}
In this work, we focus our analysis on the RS of cluster galaxies. We
investigate the development of the RS by applying a numerical
technique which combines cosmological non-radiative {\em N}-Body/SPH
(Smoothed Particle Hydrodynamics) simulations of galaxy clusters in a
concordance $\Lambda$ Cold Dark Matter universe and a semi-analytic
model of galaxy formation and evolution. In this hybrid model, the
outputs of the cosmological simulation are used to construct merger
histories of dark matter (DM) haloes and their embedded substructures,
which are then used by the semi-analytic code to generate the galaxy
population. In this Section we briefly describe the hybrid model.

\subsection{Simulated galaxy clusters}\label{Sec:SimulatedClusters}
We consider two sets of simulated galaxy clusters, C14 and C15, which
contain clusters with virial masses in the ranges $\simeq
(1.1-1.2)\times 10^{14}\,h^{-1}\,{\rm M}_\odot$ and $\simeq
(1.3-2.3)\times 10^{15}\,h^{-1}\,{\rm M}_\odot$, respectively. The
clusters in the C14 set are named ${\rm g}1542$, ${\rm g}3344$, ${\rm
  g}6212$, ${\rm g}676$ and ${\rm g}914$,whereas those in C15 are
${\rm g}1$, ${\rm g}8$, ${\rm g}51$ \citep[see][for
  details]{Dolag05}. These clusters have been initially selected from
a DM-only simulation with a box size of $480\,h^{-1}$~Mpc
\citep{Yoshida01}, for a cosmological model characterized by
$\Omega_{\rm m}$ = 0.3, $\Omega_{\Lambda}$ = 0.7, ${\rm H}_{\rm o}= 70
\,{\rm km} \, {\rm s}^{-1} \, {\rm Mpc}^{-1}$, $\Omega_{\rm b}=0.039$
for the baryon density parameter, and $\sigma_8=0.9$ for the
normalization of the power spectrum. The Lagrangian regions
surrounding the selected clusters have been re-simulated at higher
mass resolution by applying the `Zoomed Initial Condition' technique
\citep{Tormen97} and the Tree-SPH {\tt GADGET-2} code
\citep{Springel05a}.  The mass resolution is the same for all
simulations, with masses of DM and gas particles $m_{\rm dm}=1.13
\times 10^{9}\,h^{-1}\,{\rm M}_{\odot}$ and $m_{\rm gas}=1.69\times
10^{8}\,h^{-1}\,{\rm M}_{\odot}$, respectively.  As for the force
resolution, the Plummer-equivalent gravitational softening is fixed at
$\varepsilon=5 \,h^{-1}$~kpc in physical units at redshift $z\le 5$,
while it switches to comoving units at higher redshifts taking a value
of $\epsilon$~=~30~$h^{-1}$~kpc.

Although the simulations are hydrodynamical, we only use the
kinematical information provided by DM particles in order to identify
DM haloes. To this aim, we consider that DM particles have their
masses increased to its original value, since gas was introduced in
the high-resolution region by splitting each parent particle into a
gas and a DM particle. Dark haloes are first identified as virialized
particle groups by a friends-of-friends (\fof) algorithm.  The {\small
  SUBFIND} algorithm \citep{Springel01} is then applied to these
groups in order to find self-bound DM substructures, which we call
subhaloes. Subhaloes are the sites of galaxy formation, and their
merger trees are built by extracting from the simulations all
subhaloes with 10 or more bound DM particles, since smaller ones are
dynamically unstable \citep{Kauff99}.

\subsection{Semi-analytic model \sag}
The semi-analytic model \sag~used here is based on that described by
\citet{Lagos08}. The physical processes considered in this model are
the cooling of hot gas as a result of radiative losses, quiescent star
formation, and starbursts during disc instability events and galaxy
mergers. Feedback from supernova explosions and active galactic nuclei
(AGN) are also considered. The mass of hot gas available in a halo is
determined at each simulation output on which the semi-analytic model
is applied. This mass is defined as the baryonic fraction of the
virial mass of the halo minus the total mass of baryons in the form of
cold gas and stars and the  mass of black holes already present within
the halo. The model tracks the circulation of metals between the
different baryonic components, that is, hot diffuse gas, cold gas and
stars, taking into account the mass evolution of different chemical
elements \citep{Cora06}.  The version of \sag~used in this work
incorporates the more detailed calculation of disc galaxy scale radii
described in \citet{Tecce10}.

After tuning the free parameters involved in the different physical
processes considered, the model is able to reproduce several
observational constraints simultaneously, such as galaxy luminosity
functions, relations between central black hole mass and other
properties of the host bulge, quasar luminosity function, stellar mass
functions at different redshifts, and mass-metallicity,
colour-magnitude and disc size-luminosity relations (see LCP08 and
\citealt{Tecce10} for details).  

During their evolution, galaxies can acquire stellar mass by SF,
either in quiescent form by converting their cold gas into stars, or
in starbursts which can be triggered by disc instability events or
galaxy mergers. During mergers, a galaxy's stellar mass also increases
due to the stars accreted from the merging satellite. In \sag, as in
many other semi-analytic codes, only central galaxies have cooling
flows; the cold gas accreted is assumed to settle into the galactic
disc. A galaxy's mass can subsequently be reduced through recycling
due to stellar winds and supernovae explosions.  The latter also
reduce the amount of cold gas available for further SF through energy
feedback.  AGN activity also injects energy into the hot phase and
decreases the flow of cooling gas into the galaxy, thus regulating SF
as well.

The complex combination of all these processes is reflected in the
resulting metallicities and luminosities of galaxies. In the present
study, we track the increase of galaxy mass considering, separately, the
contribution of new stars formed during SF episodes of different types
(quiescent mode and starburst modes) and stars accreted from satellite
galaxies during mergers. \sag~considers both major and minor mergers, and
in this work we also identify them as either wet or dry mergers.  In the
following, we give a short description of the parameters that characterize
different types of mergers, and refer the reader to LCP08 for details on
quiescent SF and disc instability events.

\subsection{Galaxy mergers and starbursts}
\label{sec:MSB}
In a hierarchical scenario of structure formation, mergers of galaxies
are a natural consequence of the mergers of DM haloes in which they
reside, and play an important role in determining the mass and
morphology of galaxies.  The galaxy catalogue is built up by applying
the semi-analytic model to the detailed DM subhalo merger trees
extracted from the cluster simulations. After the identification of DM
substructures within \fof~ groups, the largest subhalo in a \fof~group
is assumed to host the central galaxy of the group, located at the
position of the most bound particle of the subhalo. These galaxies are
designated as {\it central} galaxies.  Central galaxies of other
smaller subhaloes contained within the same \fof~ group are referred
to as {\it halo} galaxies. The subhaloes hosting these galaxies are
still intact after falling into larger structures. There is a third
group of galaxies generated when two subhaloes merge and the galaxy of
the smaller one becomes a satellite of the remnant subhalo.  We assume
that these {\it satellite} galaxies merge with their corresponding
subhalo central galaxy on a dynamical friction time-scale.

Mergers are classified as {\em major} or {\em minor} according to the
ratio between the baryonic mass (stellar plus cold gas mass) of the
accreted satellite galaxy and of the  central galaxy,
$f_{\rm merge}=M^{\rm sat}/M^{\rm central}$. A major merger occurs
when $f_{\rm merge}>0.3$. Otherwise, we are in the presence of a
minor merger.

In the case of a major merger, all stars present in the merging
galaxies are rearranged into a spheroid. Additionally, all the cold
gas available in the remnant is consumed in a starburst as a result of
the perturbation introduced by the merging satellite. This
perturbation drives the cold gas into the bulge component of the
remnant, where it is completely transformed into stars. Thus, major
mergers lead to the creation of an elliptical galaxy.

On the other hand, during minor mergers all the stars of the merging
satellite galaxy are added to the spheroid of the remnant, whereas the
stellar disc of the accreting galaxy remains unaltered.  The presence
of a starburst during a minor merger depends on the gas mass fraction
of the disc of the central galaxy, $f_{\rm ColdGas}^{\rm
  central}=M_{\rm ColdGas}^{\rm central} /M_{\rm disc}^{\rm central}
$, where $M_{\rm disc}=M_{\rm Stellar}-M_{\rm Bulge}+ M_{\rm
  ColdGas}$, $M_{\rm Stellar}$ is the total stellar mass of the
galaxy, $M_{\rm Bulge}$ is the mass of the bulge (formed only by
stars), and $M_{\rm ColdGas}$ is the cold gas content of the
galaxy. If $f_{\rm ColdGas}^{\rm central}>f_{\rm gas,minor}$, all the
cold gas in the merging galaxies is assumed to undergo a starburst;
stars formed in this fashion are then added to the bulge of the
remnant. The inclusion of the minor merger starburst threshold
($f_{\rm gas,minor}$) is motivated by the suggestion of \citet{HM95}
that gas-rich discs should be susceptible to bursts triggered by the
accretion of small satellites. \citet{Malbon07} consider this SF
channel in a semi-analytic code; varying the minor merger starburst
threshold with respect to the value adopted by them, we find that we
are able to reproduce the observed luminosity function with $f_{\rm
  gas,minor}=0.6$ (as done in LCP08). This type of merger is
considered a {\em minor wet merger} since it leads to SF. In this
case, the stellar mass of the central galaxy is increased by the new
stars formed and by the stellar content of the accreted satellite
galaxy.

On the contrary, if $f_{\rm ColdGas}^{\rm central}<f_{\rm gas,minor}$
then no burst occurs, and we classify it as a {\em minor dry
  merger}. Furthermore, bursts do not occur if the satellite is much
less massive than the central galaxy ($f_{\rm merge}<f_{\rm burst}$,
with $f_{\rm burst}$=0.05), regardless of the amount of gas in the
central disc. Since this type of minor merger does not trigger
starbursts, it is also considered a minor dry merger.  The choice of
these parameters allows us to reconcile observational data with the
properties of the galactic populations given by \sag.

For the sake of our study, we also need to classify major mergers as
either wet or dry. This classification does not affect how \sag~works,
and is only of interest for the identification of different types of
processes that contribute to galaxy formation. Then, once we have
identified a major merger, we classify it as wet or dry depending on
the gas fraction of the disc of the remnant. Then, if $f_{\rm
  ColdGas}^{\rm remnant}<f_{\rm gas,major}$ the major merger is dry,
and it is wet otherwise.  We adopt the value $f_{\rm gas,major}=0.4$;
further discussion regarding the choice of this parameter will be
given in Section \ref{Sec:Proc}.

\begin{figure*}
\begin{center}
\includegraphics[width=0.40\textwidth]{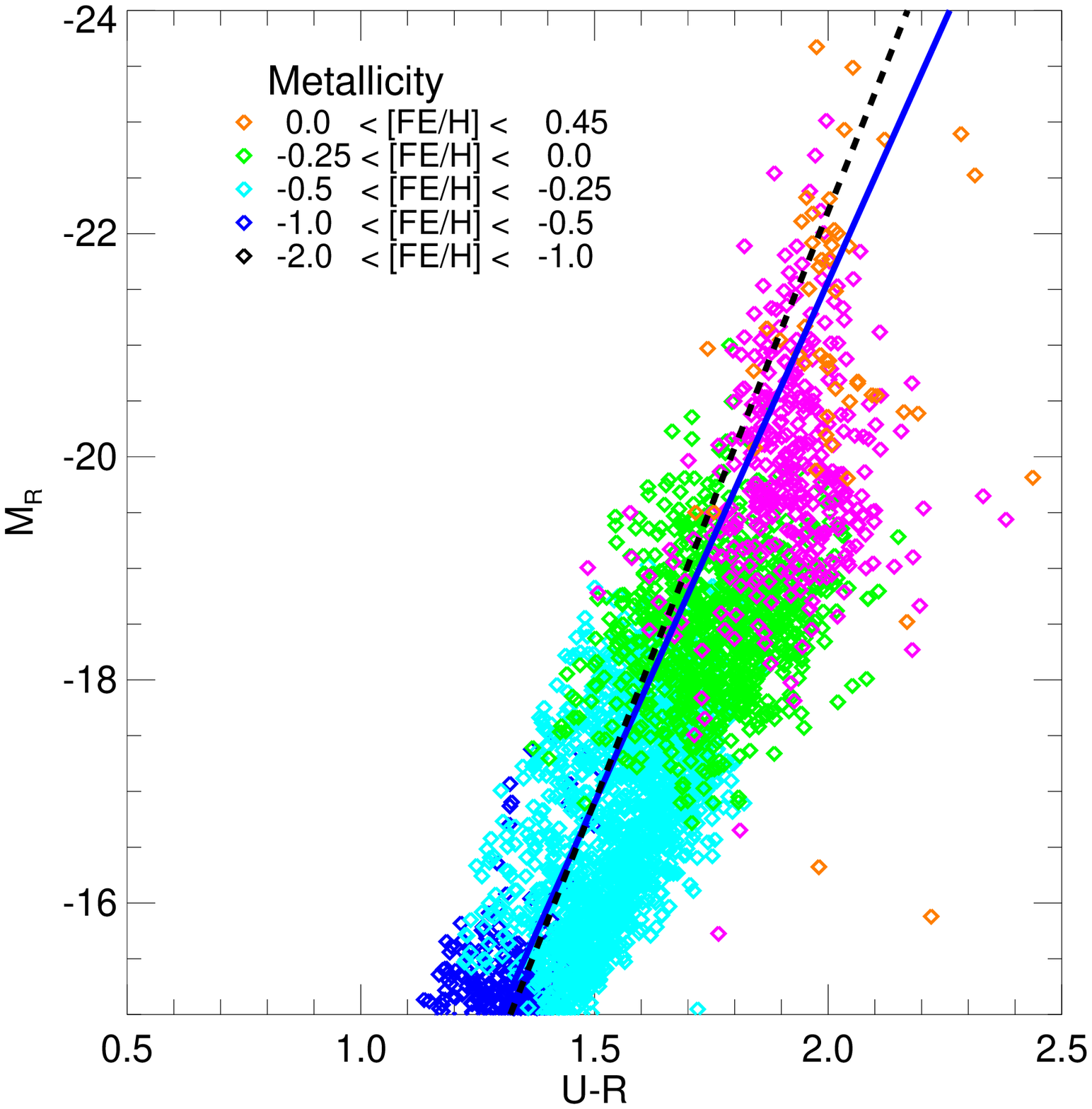}
\includegraphics[width=0.40\textwidth]{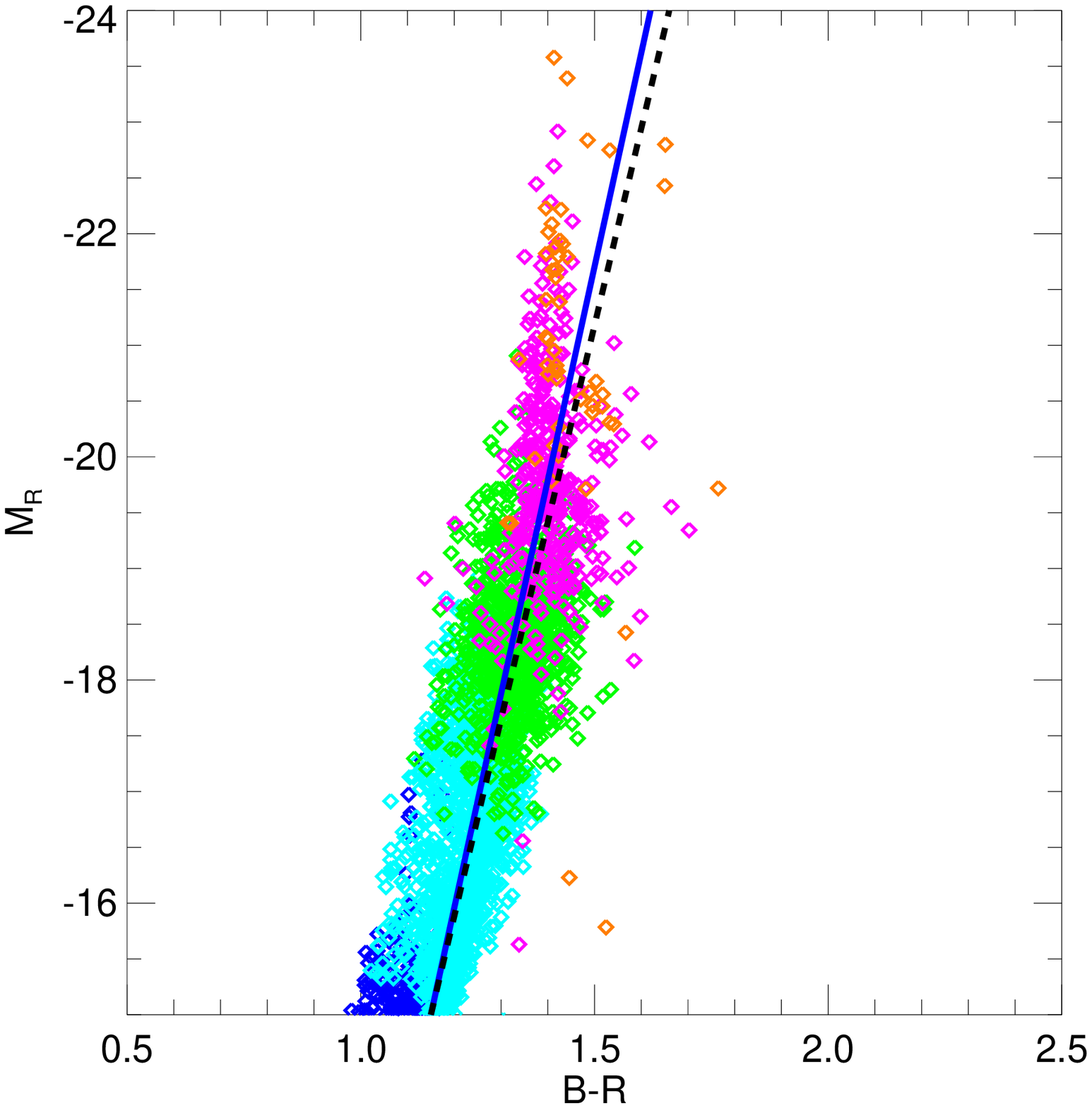}
\includegraphics[width=0.40\textwidth]{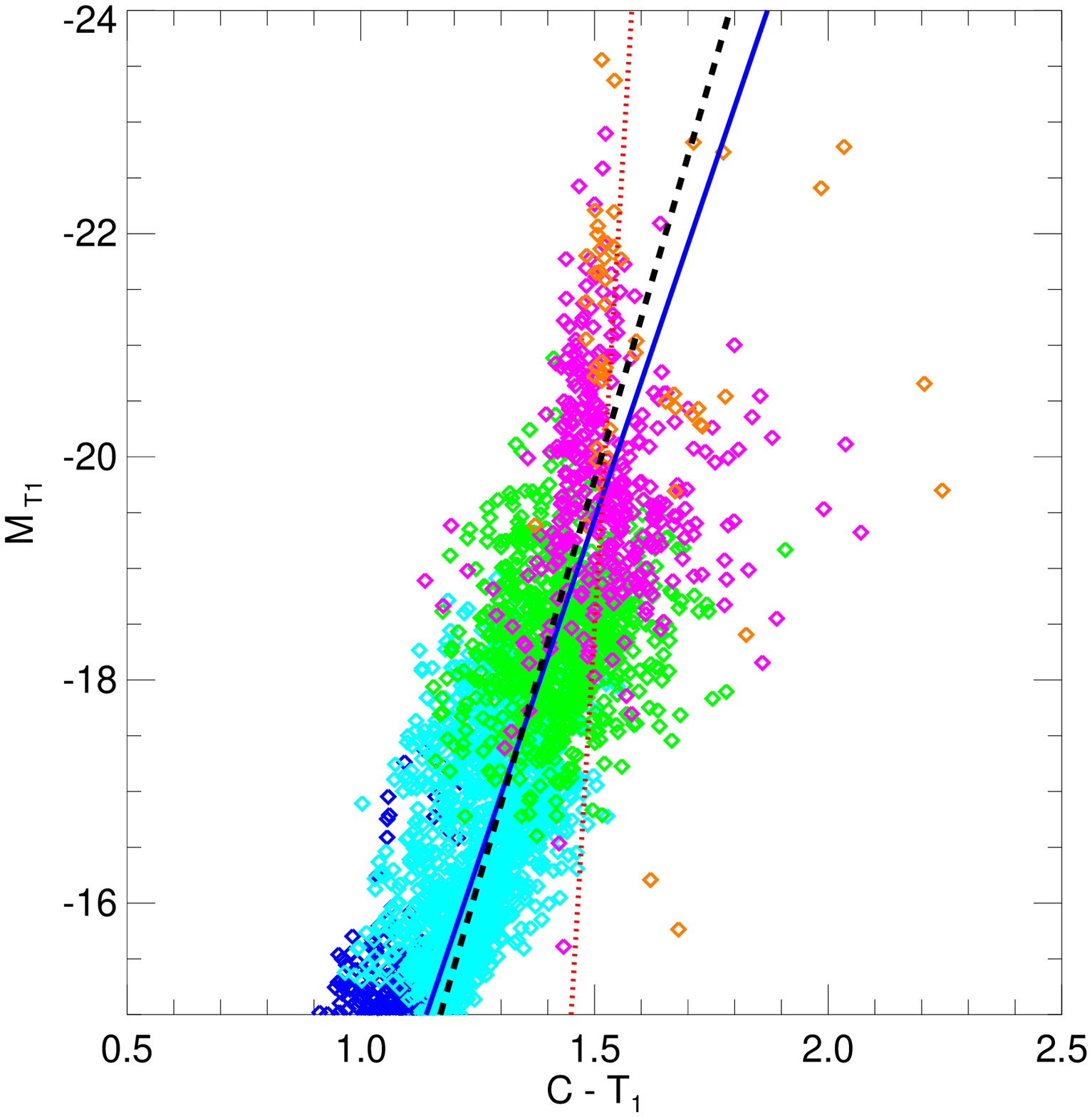}
\includegraphics[width=0.40\textwidth]{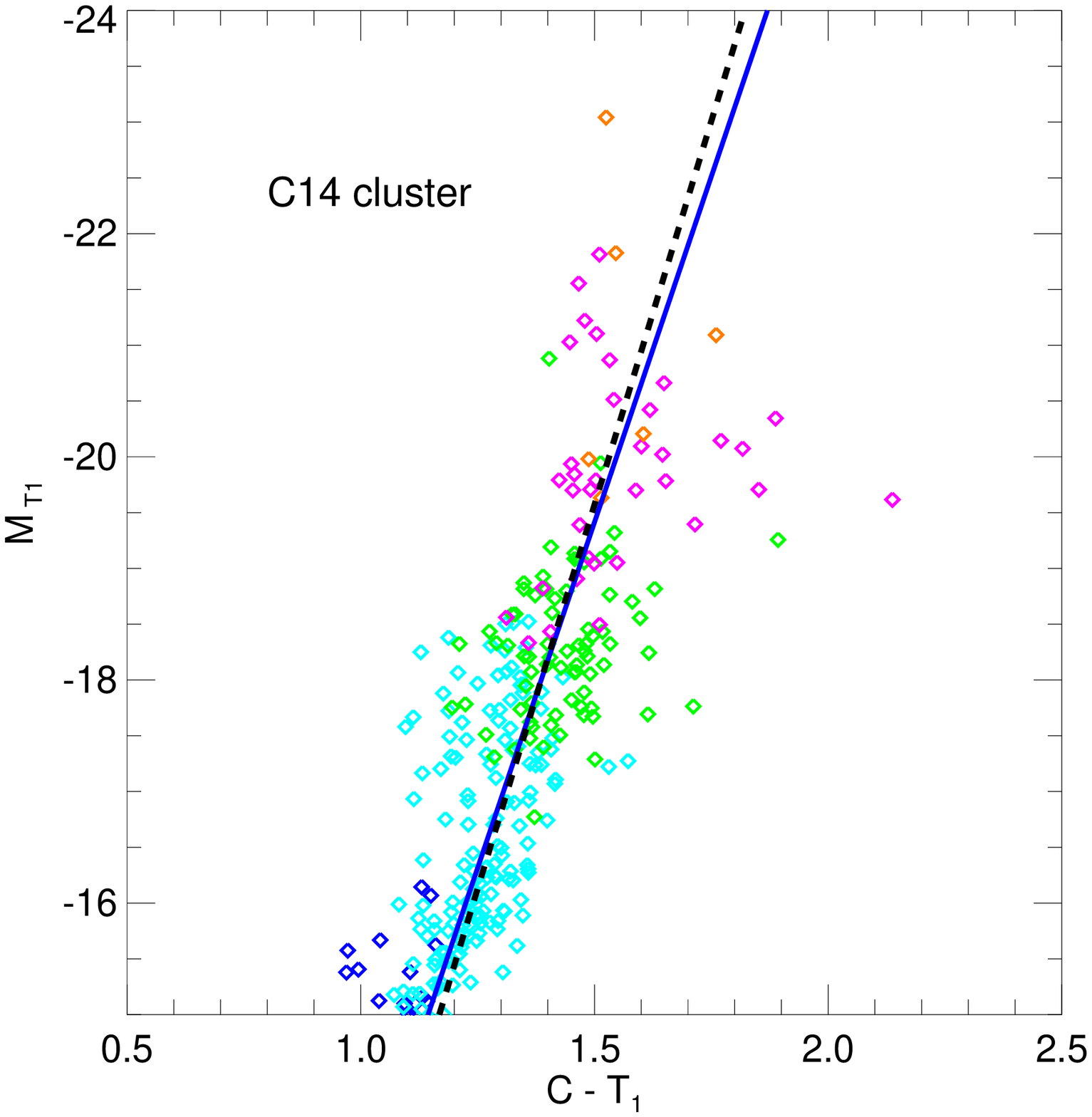}
\caption{RS obtained from the semi-analytic model \sag~applied to the
  C15$\_$3 and C14$\_$2 simulated galaxy clusters (see Table~1) in
  different photometric systems: $R$ vs. $(U-R)$ and $R$ vs. $(B-R)$
  for C15$\_$3 cluster (top), and $T_{\rm 1}$ vs. $(C-T_{\rm 1})$ for
  C15$\_$3 cluster (bottom left) and C14$\_$2 cluster (bottom
  right). Symbols are coloured according to the iron abundance of the
  stellar component of the simulated galaxies. Least-squares fits to
  the faint end of the simulated RS and to the whole relation are
  represented by blue solid lines and black dashed lines,
  respectively. The fit to the bright end of the simulated RS of the
  C15 cluster is also shown in the plane $T_{\rm 1}$ vs. $(C-T_{\rm
    1})$(red dotted line).}
\label{fig1}
\end{center}
\end{figure*}

\section{Colour-magnitude relation}\label{Sec:RedSeq}  
On the basis of the results obtained by applying our semi-analytic
code \sag~to the sets of simulated galaxy clusters C14 and C15, we are
able to construct the corresponding RS, whose bright end is mostly
populated by ETGs. In the simulations, galaxies belonging to the RS
are selected according to the empirical redshift-dependent separation
between colour sequences given by \citet{Bell04}, based on the bimodal
distribution of observed galaxy colours out to $z\approx 1$. In this
way, galaxies redder than 
\begin{equation}
 \langle U-V \rangle = 1.15 -0.31\,z- 0.08 \,(M_{\rm V} -5 \rm{log  h}
 +20),
\end{equation}
for $0 < z < 1$, are considered as belonging to the RS. In the
following, we adopt $z = 0$ to obtain simulated RS at the present
epoch and compare them with observed CMRs of ETGs.  As we will be
basically dealing with the CMR slope but not with its zero-point, we
do not need to convert observational data from absolute into apparent
magnitudes and vice-versa. In the model, neither magnitudes are
influenced by extinction nor the colours are affected by reddening.

\subsection{Slopes of the colour-magnitude relations in different bands}
Our semi-amalytic model provides the magnitudes of galaxies in the
Johnson-Morgan system.  In order to compare with observed data, they
have to be transformed to other photometric systems.  Fig. \ref{fig1}
shows the RS obtained for one of the more massive simulated clusters
(C15$\_$3) in three different colour-magnitude planes: $R$ vs. $U-R$,
$R$ vs. $(B-R)$, and $T_{\rm 1}$ vs. $(C-T_{\rm 1})$. We also show the
RS of a less massive cluster (C14$\_2$) in the $T_{\rm 1}$
vs. $(C-T_{\rm 1})$ plane in the right bottom panel. The $R$ magnitude
is in the Cousins system, while the $C$ and $T_{\rm 1}$ magnitudes
correspond to the Washington photometric system. The $(C-T_{\rm 1})$
integrated colour has proved to be quite sensitive to metallicity for
old, single stellar populations like globular clusters (e.g., see
\citealt{Harris2002, Forte2007}), while the $(U-R)$ integrated colour
is age sensitive. Besides, the RS in the $V$ vs. ($V-I$) plane is also
considered in the analysis, but not shown in this figure.

The $R$ and $I$ magnitudes in the Johnson-Morgan system are
transformed into the Cousins system through the relations given by
\citet{Fukugita95} for ETGs. The photometric conversion from
Johnson-Morgan $V$ minus Cousins $I$ colours into the Washington
$(C-T_{\rm 1})$ ones is performed through the transformation $ (V-I)=
0.49(C-T_{\rm 1}) + 0.32$, given by  \citet{ForbesForte01} for
globular clusters, under the assumption that ETGs are mainly old
stellar systems as well. Finally, $T_{\rm 1}$ magnitudes are obtained
from the Cousins $R$ ones by applying the relation $R-T_{\rm 1}\sim
-0.02$, given by \citet{Geisler96}.

It can be clearly seen from Fig. \ref{fig1} that the simulated RS of
the C15 cluster, in the three colour-magnitude planes, present a break
with brighter galaxies detached from the general trend denoted by the
fainter ones. This break occurs at approximately the same magnitude in
the different systems ($M_R^{\rm break} \sim M_V^{\rm break} \sim
M_{T_1}^{\rm break} \approx -20$; galaxy mass of $\sim 10^{10}
M_{\odot}$), being more evident in the Washington system, with
brighter galaxies displaying an almost constant colour ($C-T_{\rm
  1}\approx 1.5$). Taking such break as a reference, we will refer
hereafter to the galaxies brighter/fainter than the magnitude of the
break as the `bright/faint end' of the RS.

\begin{table*}
  \caption{Slopes calculated from least-squares fits to both the faint
    end of the RS and the whole relation, for the different simulated
    clusters in several photometric bands.}
  \begin{center}
    \begin{tabular}{l|c c|c c|c c|c c}
      & \multicolumn{2}{|c|}{$R$ vs. $B-R$}& \multicolumn{2}{|c|}{$V$  vs. $V-I$}&  \multicolumn{2}{|c}{$T_1$ vs. $C-T_1$}& \multicolumn{2}{|c}{$R$ vs. $U-R$}\\ \hline
      Cluster & Faint end &  All  & Faint end  & All & Faint end & All   & Faint end & All \\\hline
      
      C14$\_$1 (g1542) &-16.92 &-19.90  &-20.12 &-25.22  &-10.79 &-12.29   &-8.55    &-9.85 \\
      C14$\_$2 (g3344) &-18.55 &-20.00  &-19.97 &-27.85  &-12.24 &-13.77   &-12.24   & -13.77\\ 
      C14$\_$3 (g6212) &-19.23 &-20.88  &-24.65 &-30.37  &-12.36 &-13.80   &-9.58    &-10.07\\ 
      C14$\_$4 (g676) &-17.43 &-20.38	&-26.47 &-28.95  &-11.45 &-13.83   &-8.75    &-9.79\\  
      C14$\_$5 (g914) &-16.39 &-18.99	&-21.48 &-25.31  &-10.47 &-12.35   &-8.30    &-9.25\\  
      \hline                                                             
      C15$\_$1 (g1) &-18.10 &-21.36	&-24.78 &-29.72  &-11.48 &-13.96   &-9.00    &-10.32\\    
      C15$\_$2 (g8) &-18.11 &-20.69	&-24.12 &-28.84  &-11.55 &-13.71   &-9.15    &-10.21\\    
      C15$\_$3 (g51) &-18.96 &-21.78  &-25.57 &-31.19  &-12.38 &-14.62     &-9.33    &-10.48\\   
      \hline      
    \end{tabular}
  \end{center}
  \label{table1}
\end{table*}

Two fits are performed to the simulated RS in each photometric system,
considering the whole relation and the faint end only. These fits are
shown in Fig. \ref{fig1} for the C15$\_$3 cluster and for C14$\_$2
cluster as black dashed and blue solid lines, respectively.  Table
\ref{table1} presents the slopes of these least-squares fits performed
to the RS of the eight simulated clusters described in
Subsection~\ref{Sec:SimulatedClusters}. Note that the slopes depicted
in Table \ref{table1}, designated as $b$ in the following analysis,
correspond to RS where the colour is the independent variable. In
order to compare with those obtained from observed CMRs, we will have
to invert the `observed' slopes  ($1/b$) in the cases that CMRs are
calculated taking the colour as the dependent variable (i.e. as if the
diagrams displayed in Fig. \ref{fig1} were rotated $90 \degr$). 

With regard to the observed CMRs, a wide cluster sample is presented
by \citet{LopezCruz04}, who study the CMRs of 57 low-redshift cluster
galaxies, with different richness, cluster types and X-ray
luminosities. The fits to the CMRs are performed in the $(B-R)$
vs. $R$ plane (see their table $1$), from which we calculate an
average slope $1/b_{\rm BR}= -0.051 \pm 0.002$, or $b_{\rm BR} =
-19.61$ to compare with our simulations. This falls within the range
of values presented in Table~\ref{table1} for the corresponding `all
CMR' fits.

\citet{Misgeld08} fit the CMR of the early-type dwarf galaxy sample
($M_{\rm V} > -17$) of Hydra\,I cluster in the $(V-I)$ vs. $V$ plane,
and get a slope $1/b_{\rm VI} = -0.039$ (rms error of the fit equal to
$0.12$), or $b_{\rm VI} = -25.64$.  They also estimate the slope of
the CMR for Local Group dwarf ellipticals (dE) and dwarf spheroidals
(dSph) in the same colour-magnitude plane as $1/b_{\rm VI} = -0.038$
(rms of the fit $0.09$), or $b_{\rm VI} = -26.32$. Still in the same
plane, a slope $1/b_{\rm VI} = -0.033 \pm 0.004$, or $b_{\rm VI} =
-30.30$, is given by \citet{Mieske07} for the dE galaxy sample
($M_{\rm V} > -17$) in the Fornax cluster. \citet{Misgeld09} analyse
the CMR followed by dwarf ETGs in the Centaurus cluster and obtain
$1/b_{\rm VI} = -0.042$ (rms of the fit $0.10$), or $b_{\rm VI} =
-23.81$.  Except for the slightly lower value from Mieske et al., all
the slopes are within the range covered by those of the `faint end'
fits listed in Table \ref{table1}.

The Perseus cluster is studied by \citet{deRijcke09} with HST/ACS
data; they present a global CMR that includes several other groups and
clusters, obtaining a common fit with slope $1/b_{\rm VI} = -0.033 \pm
0.004$, or $b_{\rm VI} = -30.30$, which is also in agreement with `all
CMR' values from Table \ref{table1}.

\citet{Analia08} obtained the CMR of ETGs located in the central
region of the Antlia cluster, using the Washington photometric
system. This CMR is characterised by the least-squares fit $T_{\rm 1}=
(38.4 \pm 1.8) -(13.6 \pm 1.0) \,(C-T_{\rm 1})$. This latter fit
corresponds to the whole galaxy range, as a break is not
distinguishable in this observed CMR. Note that this is so because the
brightest galaxies in this sample hardly reach the magnitudes above
the break detected in the simulations.

The linear fit to the bright end of the RS in this colour-magnitude
plane for the C15 cluster is represented in the bottom left panel of
Fig. \ref{fig1} by a red dotted line, and gives $M_{T_{\rm 1}}= 83.65
- 67.92 (C-T_{\rm 1})$. Its slope clearly differs from the one
corresponding to the faint end, which is fit by the relation $M
_{T_{\rm 1}}= 0.82 -12.38 (C-T_{\rm 1})$, shown with a blue solid
line. This latter fit is closer to the one applied to the whole RS,
for which we obtain $M_{T_{\rm 1}}= 2.19 - 14.62 (C-T_{\rm 1})$. The
slope of the fit to the whole RS of the C14$\_2$ cluster is $-13.77$
(see Table \ref{table1}) in very good agreement with the observed
value  $-13.6$ of the slope for Antlia.  Similar results have been
found in all the simulated clusters, as can be seen from
Table~\ref{table1}. This supports the idea of the universality of the
CMR as inferred from observational results
\citep[e.g][]{LopezCruz04,deRijcke09}.

The detachment of the bright end ($-24 \lesssim M_{T_{\rm 1}}\lesssim
-20$) with respect to the linear fit to the faint end is clear for the
three massive simulated clusters (C15), although the change of slope
in the RS of the less massive clusters (C14) is not so evident. This
is mainly due to the lack of objects in the high luminosity range of
C14 clusters and the intrinsic scatter of the relation. Only two small
groups ($\rm{C14\_1, C14\_2}$) show a slightly tighter relation at
$M_{T_{1}} \leq -20 $, where the change of slope becomes
perceptible. As mentioned above, this effect is present in some
observed CMRs as well. For instance, it can be seen in the CMR of
Hydra\,I cluster presented by \citet{Misgeld08} (their Figs.~2 and
10), though it is not particularly noticed by the authors. The studies
of the Virgo cluster by \citet{Janz09} and of a large sample of local
ETGs by \citet{Skelton09}, both based on SDSS data, clearly show a
tilt towards bluer colours at the bright end of their CMRs.

We find a very good agreement between the slopes of the observed CMR
and the simulated RS, specially for galaxies fainter than the break
$M_{T_1}^{\rm break} \sim -20$. It is remarkable that this global
agreement, as well as the tilt of the bright end towards bluer
colours, emerge naturally from \sag~after calibrating the code to
satisfy several other observational constraints simultaneously, as
mentioned in Section~\ref{Sec:model}.

\begin{figure}
\begin{center}
\includegraphics[width=0.40\textwidth]{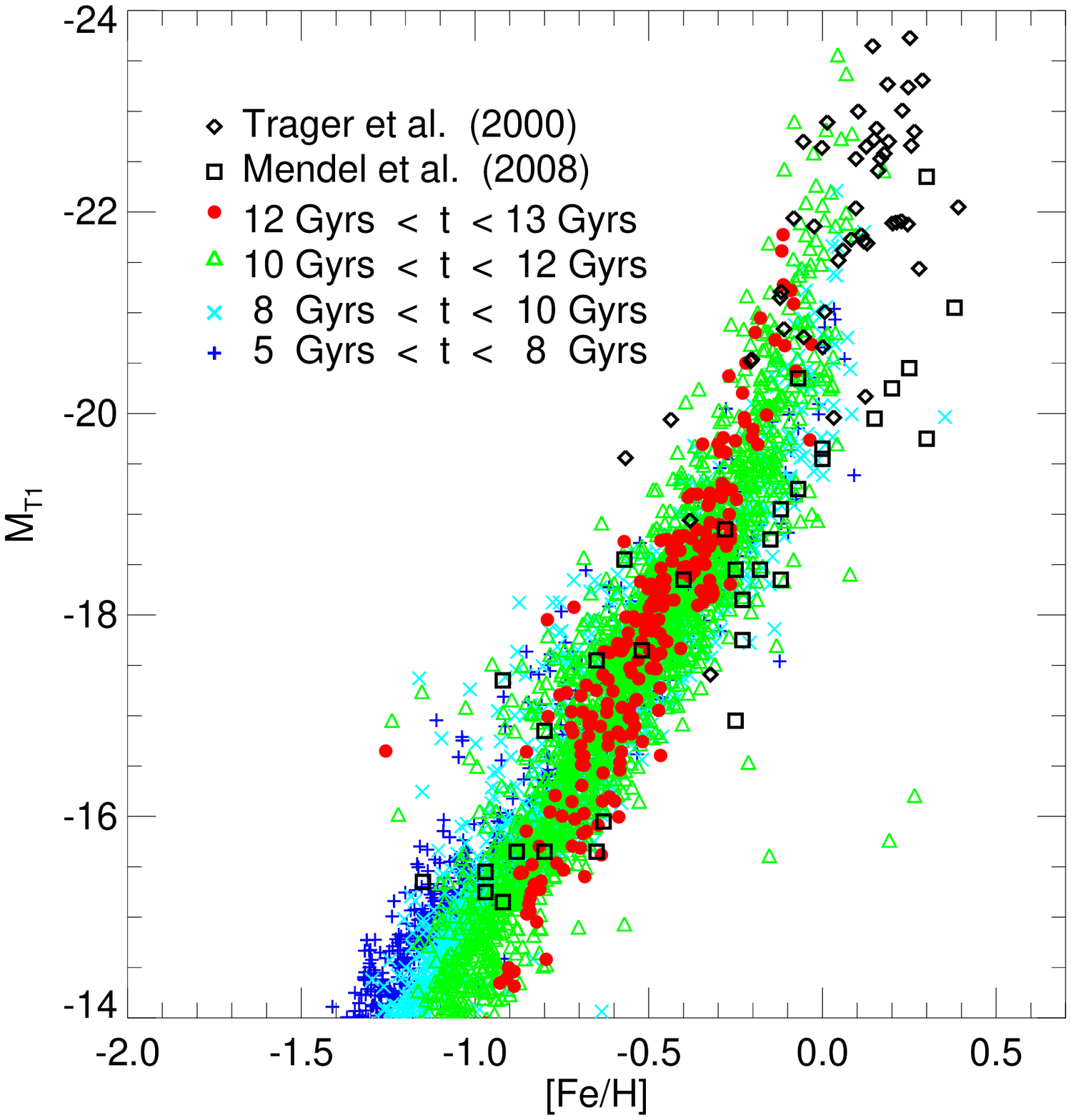}
\includegraphics[width=0.40\textwidth]{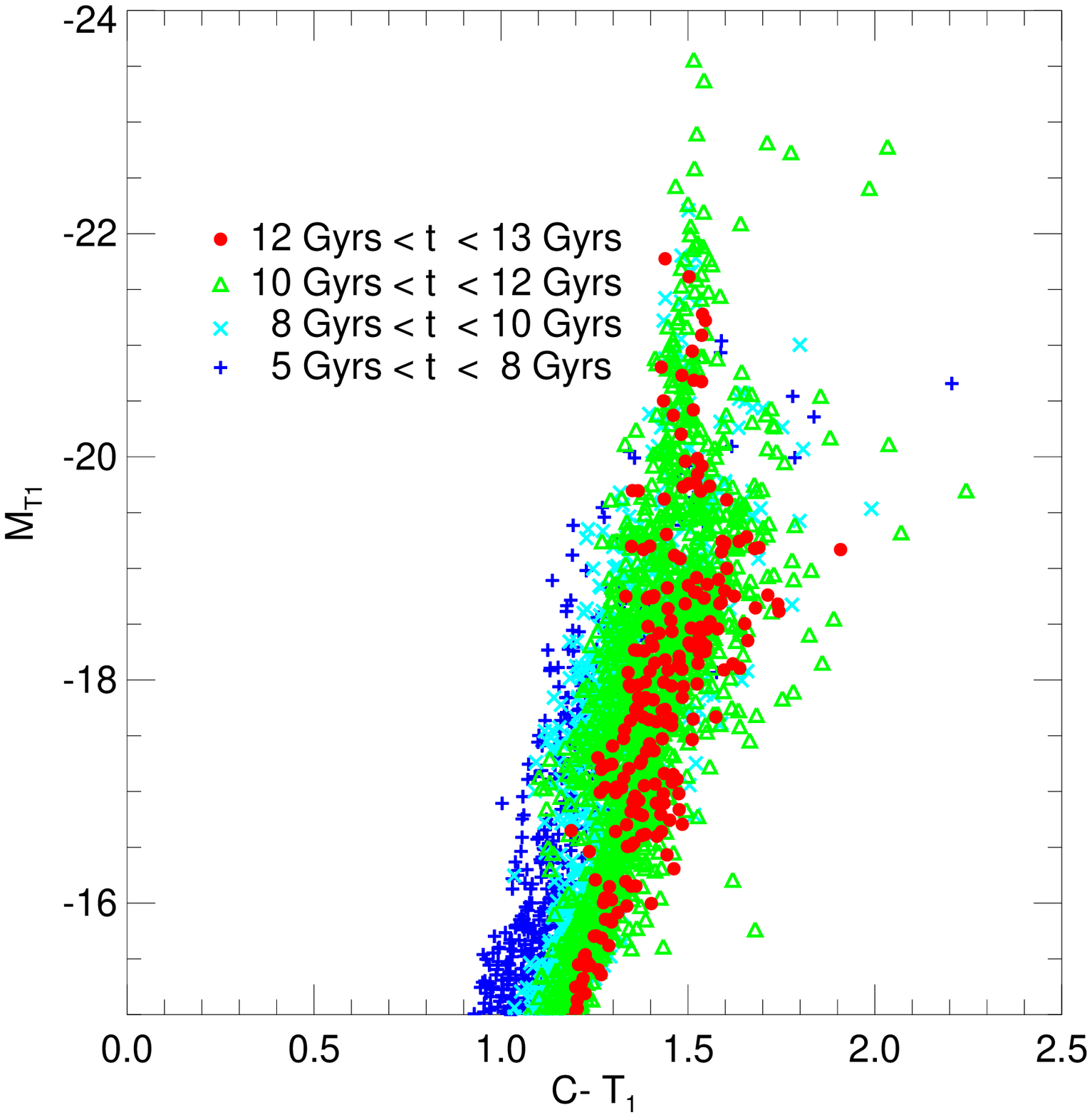}
\caption{Top panel: Luminosity-metallicity relation for the same
  simulated galaxies in the RS of the  C15$\_$3 cluster shown in
  Fig.~\ref{fig1}. Differently coloured symbols identify galaxies
  within different ranges of age; black open diamonds represent the
  [Fe/H] abundances of early-type galaxies belonging to the sample
  described in \citet{Trager00}, whereas black open squares correspond
  to galaxies in the NGC\,5044 group \citep{Mendel09}. Bottom panel:
  RS of simulated galaxies in the magnitude plane $T_{\rm 1}$
  vs. $(C-T_{\rm 1})$ colour-coded according to the range of ages in
  which they lie.}
\label{fig2}
\end{center}
\end{figure}

\subsection{Luminosity-metallicity relation}
One of the main properties also affected by the physical processes
driving the development of the CMR is the galaxy
metallicity. Simulated galaxies that form the RS shown in
Fig. \ref{fig1} are identified by different coloured symbols according
to the metallicity range in which they lie. The metallicity of each
galaxy is characterised by the iron abundance of its stellar component
by using the index $\rm{[Fe/H]}$, where we are considering the solar
value of iron abundance by number $\rm{(Fe/H)}_\odot=2.82 \times
10^{-5}$ \citep{Asplund05}. We can see that galaxies at the bright end
are the most metal rich ones with $\rm{[Fe/H]} \gtrsim -0.25$, some of
them reaching values as high as $\rm{[Fe/H]} \approx 0.43$; the mean
value of their metallicity distribution is $\langle
\rm{[Fe/H]}\rangle= -0.03 \pm 0.05$.  As we move down along the RS,
galaxies become fainter, bluer and less chemically enriched. This plot
clearly reflects the fact that the RS can be interpreted as a
mass-metallicity relation.

We can further test whether a luminosity-metallicity relation is
present. In the upper panel of Fig. \ref{fig2}, we show this relation
for the same simulated galaxies that form the RS of the C15$\_$3
cluster shown in Fig. \ref{fig1}. In order to compare our results with
observed data, we include [Fe/H] abundances of early-type galaxies
belonging to the NGC\,5044 group \citep{Mendel09}, and of a sample
described in \citet{Trager00}. The latter contains cluster members
(six of the Virgo cluster, eleven of the Fornax clusters, one of the
rich cluster Abell 194), but most of the galaxies are in small groups
of varying richness. The metallicity of this set of galaxies were
re-determined by \citet{Trager08}; [Fe/H] abundances are then obtained
from these metallicity values using a transformation given by
\citet{Trager00} (see their Eq.~1). In all cases, we have transformed
the absolute magnitudes to the $T_1$ band through the relations of
\citet{Fukugita95}. As can be seen, the observed values, which cover a
wide range of magnitudes and abundances, overlap the simulated
relation. The agreement is particularly good for galaxies in the faint
end of the RS  ($M_{T_{\rm 1}} \gtrsim -20$), although we note that
the observed abundances of some galaxies in the bright end are
slightly larger than those obtained from our model. This discrepancy
may be due to the fact that most of the galaxies in the observed
sample are in relatively low-density environments, having different SF
and enrichment histories than those residing in massive clusters.The
fact that the metallicities of galaxies at $z = 0$ satisfy
observational constraints is particularly relevant for our study. It
supports the use of the chemical history of galaxies as a tool to help
understand the development of the RS and its special feature at the
bright end.

Galaxies in the luminosity-metallicity relation of Fig. \ref{fig2}
have been colour-coded according to the range of ages in which they
lie.  Ages have been estimated considering the stellar mass weighted
mean of the birth-time of each single stellar population.  A clear
correlation between age and metallicity is evident for the least
luminous galaxies ($M_{\rm T_1} \gtrsim -16$).  Younger galaxies are
more metal poor, lying on the left side of the luminosity-metallicity
relation, while older galaxies have higher iron abundances being
located towards the right.  As we move to brighter galaxies along this
relation, ages and metallicity become anticorrelated, consistent with
the trend found by \citet{Gallazzi06} for galaxies in the SDSS at
fixed velocity dispersion.  This anticorrelation might explain the
modest scatter of the bright end of the CMR \citep{Trager00}, which is
characterized by a negligible spread in age, also consistent with
observational results by \citet{Gallazzi06}.  Most galaxies in the
bright end of our simulated RS share very similar ages ($1.0\times
10^{10}$ yr $<t<1.2\times 10^{10}$ yr).  This can be seen in the lower
panel of Fig. \ref{fig2}, where galaxies along the RS in the $T_{\rm
  1}$ vs. $(C-T_{\rm 1})$ plane are identified with different coloured
symbols according to their age.  From this figure, it is clear that,
in the faint end, the scatter in colour about the RS for a fixed
luminosity is caused entirely by age variations, as assumed by
\citet{Bernardi05}, where younger galaxies correlate with bluer
colours.  As we move to higher luminosities, galaxies of a given age
become redder because they are more metal rich. The effect of age
differences on the final colours of galaxies in the bright end of the
RS is completely negligible (see fig. 1 of \citealt{Bruzual03}), and
we can attribute to metallicity effects the behaviour of the galaxies
in the bright-end of the RS. In the following, we analyze the physical
reasons that make the old galaxies in the bright end reach an upper
limit in metallicity. Our aim is to demonstrate that this metallicity
effect is the responsible of the colours achieved by massive galaxies
which are bluer than those that would be obtained from an
extrapolation of the faint end of the RS.

\section{Physical processes involved in the development of the RS}
\label{Sec:Proc}
The galactic properties that determine the RS, that is, masses and
metallicities, are the result of a complex combination of different
physical processes involved in the formation and evolution of
galaxies. Thus, in order to identify which mechanisms are responsible
for the different features of the RS, and in particular the break at
the bright end, we track the evolution of the masses and metallicities
of the stars added to each galaxy by different processes: quiescent
star formation from the cold gas available in the galaxy disc,
starbursts during mergers and disc instability events, and the stellar
mass accreted from satellite galaxies during mergers.

In the following, the stellar mass generated during any type of star
formation event is denoted as `stars', while the stellar mass accreted
during mergers, already present in the satellite galaxy, is referred
to as `sat'.  Thus, `stars' and `sat' indicate the class of stellar
`component' contributed by a given `process'.  Hence, the stellar mass
of a galaxy $g$ at a certain redshift $z$ that results from the
addition of a certain stellar component is denoted by $SM_{g,z}^{\rm
  proc,\ comp}$.

Therefore, we consider a set of variables that take into account the
different contributions of stellar mass to a given galaxy, according
to the process involved. Specifically, 

\begin{enumerate}
\item stellar mass from quiescent star formation:

\begin{itemize}
\item $SM_{g,z}^{\rm quiescent,\ stars}$
\end{itemize}

\item stellar mass from disc instability events: 

\begin{itemize}
\item $SM_{g,z}^{\rm disc,\ stars}$
\end{itemize}

\item stellar mass from merger events:

\begin{itemize}
\item    $SM_{g,z}^{\rm minor\  dry,\ sat}$
\item    $SM_{g,z}^{\rm minor\  wet,\ sat}$
\item    $SM_{g,z}^{\rm minor\  wet,\ stars}$
\item    $SM_{g,z}^{\rm major\  dry,\ sat}$
\item    $SM_{g,z}^{\rm major\  dry,\ stars}$
\item    $SM_{g,z}^{\rm major\  wet,\ sat}$
\item    $SM_{g,z}^{\rm major\  wet,\ stars}$
\end{itemize}

\end{enumerate}

\begin{figure}
\begin{center}
  \includegraphics[width=0.45\textwidth]{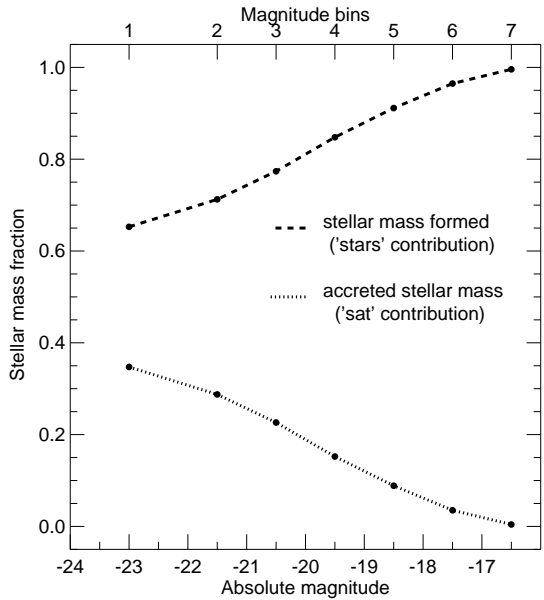}
  \caption{Stellar mass fractions of `stars' and `sat' contributions
    reached at $z=0$ by galaxies in the C15 clusters for the different
    magnitude bins.}
\label{fig3}
\end{center}
\end{figure}

\noindent With all these separate variables, we track the mass
assembly history of galaxies and the evolution with redshift of the
stellar mass fractions contributed by the different processes. Similar
variables are defined to follow the evolution of the metallicities of
galaxies, given by the ratio ${\rm Fe/H}$.  For each process
considered, we keep track of the mass of Fe and H received by a galaxy
from formed and/or accreted stars at a given redshift as a result of a
certain process; thus, we have the quantities ${\rm Fe}_{g,z}^{\rm
  proc \, comp}$ and ${\rm H}_{g,z}^{\rm proc \, comp}$, respectively.
In order to achieve a better visualization of the mechanisms that
determine the properties of galaxies along the RS, we divide the
simulated RS into six bins of one magnitude from $M_{T_{1}}=-16$ to
$M_{T_{1}} =-22$, and a seventh one from $M_{T_{1}}=-22$ to $M_{T_{1}}
=-24$. This last bin comprises two magnitudes to improve the statistic
analysis, in view of the low number of very luminous galaxies in all
simulated clusters. Thus, bin $=1$ and bin $=7$ correspond to the
magnitude ranges $-24 < M_{T_{1}} \leq -22$ and $-17 < M_{T_{1}} \leq
-16$, respectively. We analyze the evolution with redshift of the
fractions of stellar mass contributed by different processes to
galaxies in a given magnitude bin, and of their mean metallicity.  To
this aim, we define several quantities that are included in these
calculations.

The total mass of stars acquired by galaxies in a magnitude bin $b$ at
redshift $z$ arising from a given component contributed by a
particular process is:
\begin{equation}
  SM_{b,z} ^{\rm proc, \,comp}= \sum_{g=1}^{N_{b,z}} SM_{g,z}^{\rm
    proc, \,comp},
\end{equation}
where $N_{b,z}$ is the number of galaxies within each bin $b$ at each
snapshot of the simulation characterized by a redshift $z$. Then,
considering the combined contribution of the two stellar components
(`stars' and `sat') by any given process, we have 
\begin{equation}
  SM_{b,z} ^{\rm proc}= SM_{b,z}^{\rm proc, \,stars}+ SM_{b,z}^{\rm proc, \,sat}.
\end{equation}
Finally, the total stellar mass acquired by all galaxies in a given
magnitude bin that results from the  contribution of all the involved
processes is 
\begin{equation}
  SM_{b,z}^{\rm{(i)+(ii)+(iii)}}= \sum_{\rm proc=1}^{N_{\rm proc}} {SM_{b,z}^{\rm proc}},
\end{equation}
where $N_{\rm proc}$ is the total number of processes considered
grouped in the sets (i), (ii) and (iii), as described previously.

In the following, all results shown correspond to averages over the
three C15 clusters; error bars are not included in the plots because
they are extremely small. We have also evaluated the situation
considering C14 clusters. 

\begin{figure*}
  \begin{center}
    \includegraphics[width=0.33\textwidth]{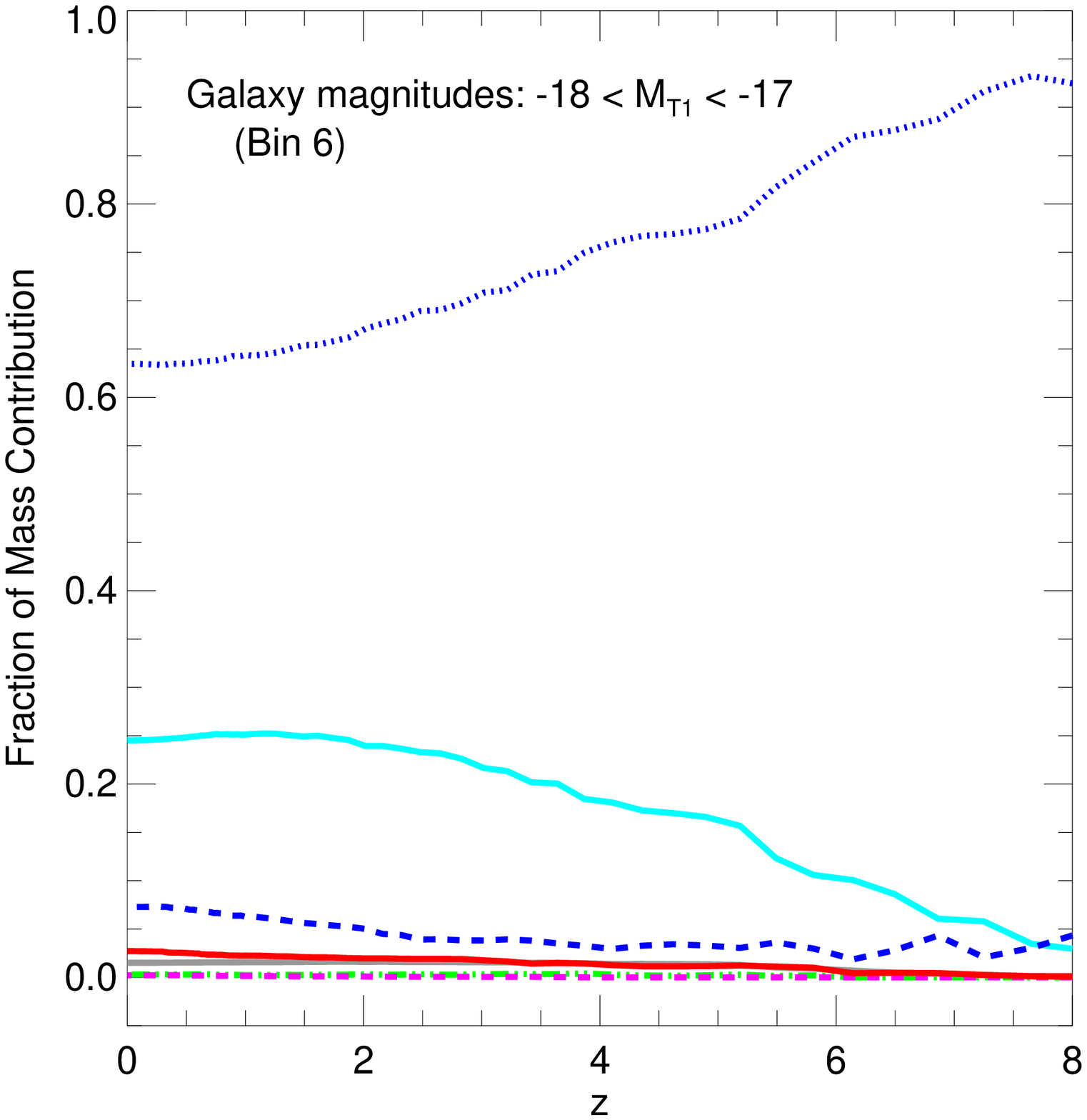}
    \includegraphics[width=0.33\textwidth]{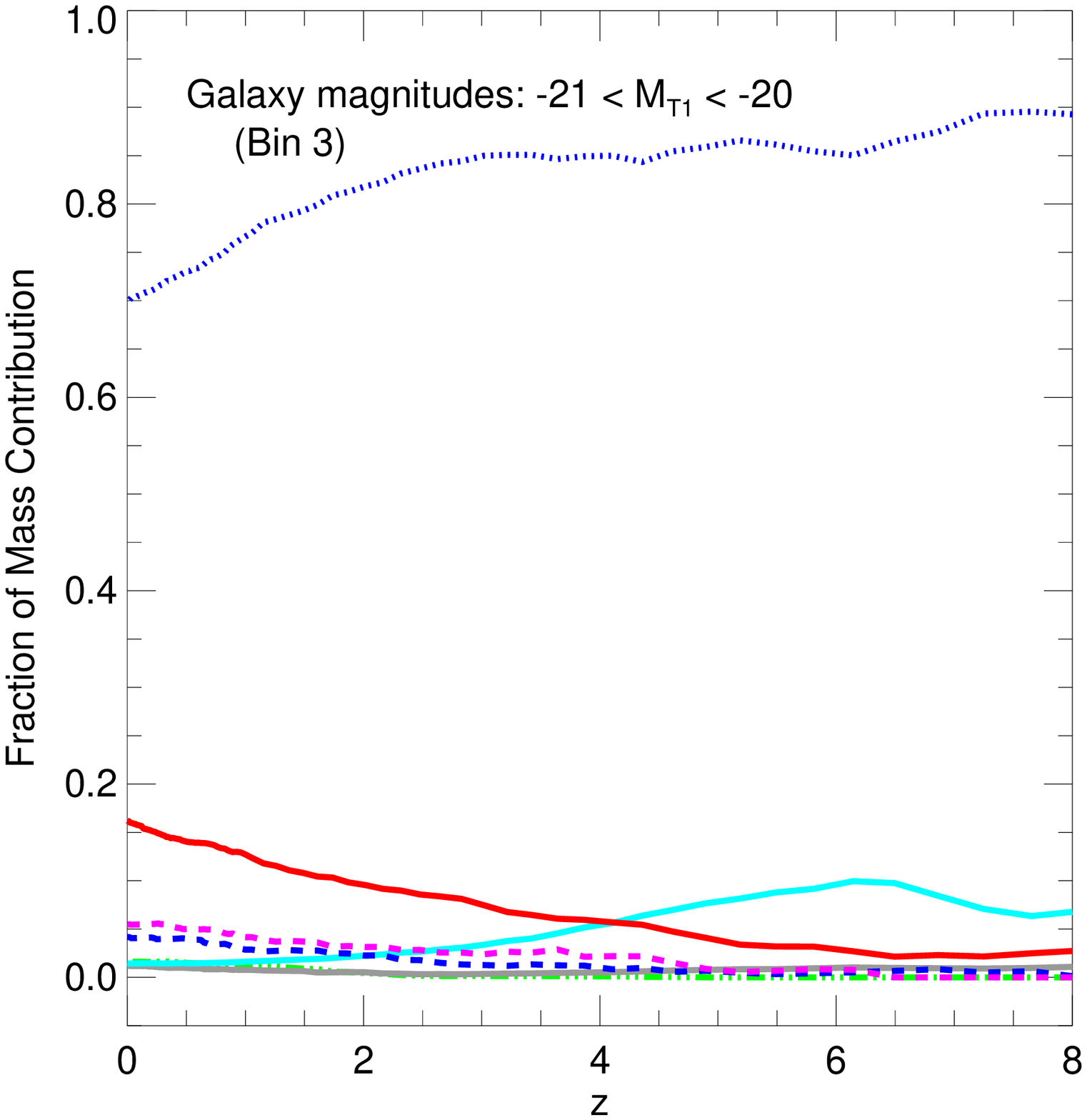}
    \includegraphics[width=0.33\textwidth]{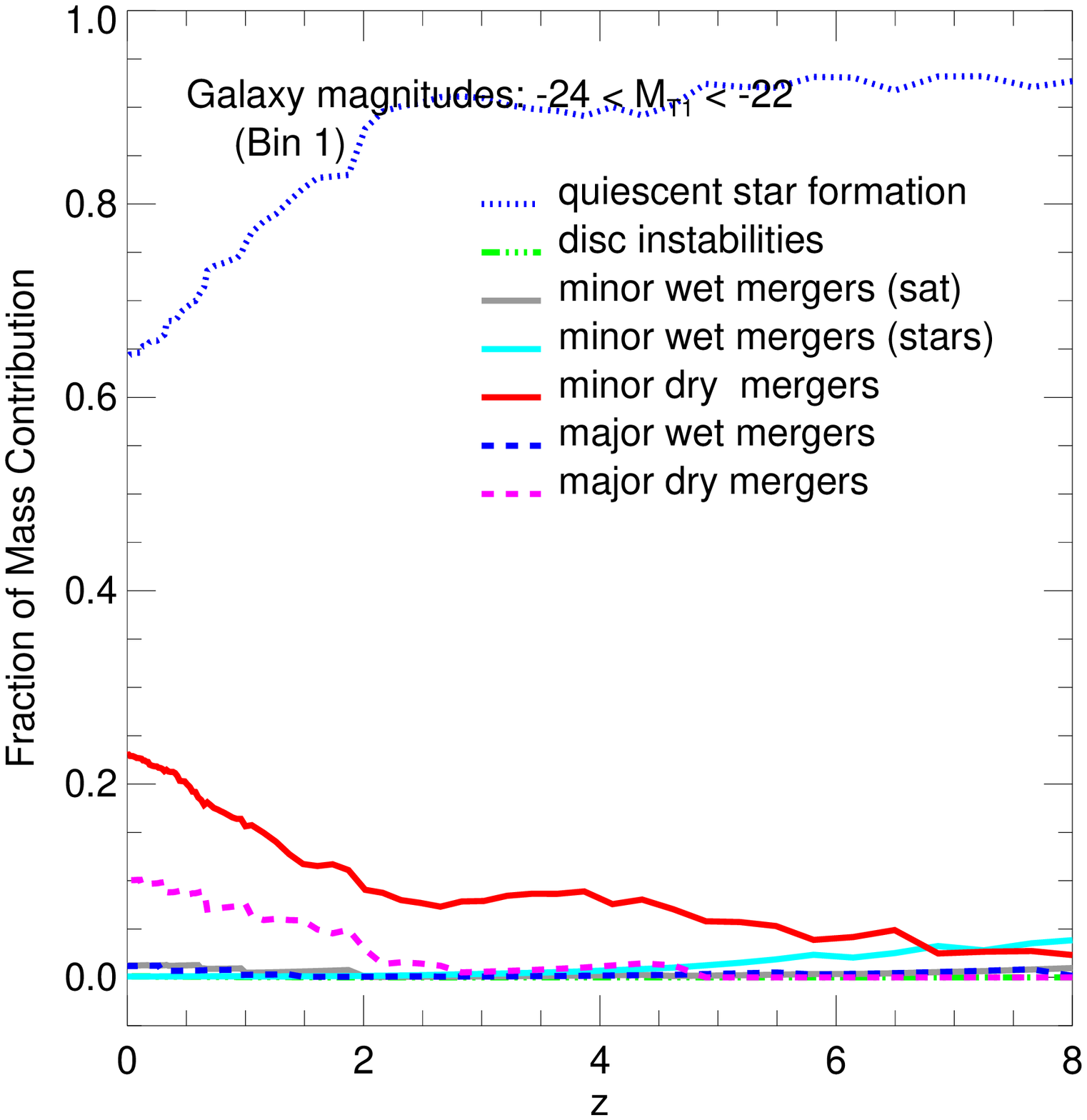}
    \caption{Evolution of the mean stellar mass fractions contributed
      by different processes for galaxies within magnitude bins $1$,
      $3$ and $6$, normalized to the total contribution to the stellar
      mass at each redshift the three C15 clusters are considered.
      Left panel: galaxies with manitudes within the range $-18 <
      M_{T_{1}}\leq -17$, magnitude bin $6$. Middle panel: galaxies
      with manitudes within the range $-21 < M_{T_{1}}\leq -20$,
      magnitude bin $3$ (critical bin). Right panel: galaxies with
      magnitudes within the range $-24 < M_{T_{1}}\leq -22$, magnitude
      bin $1$.}
    \label{fig4}
  \end{center}
\end{figure*}

\subsection{Evolution of stellar mass fractions}\label{MassFrac}
Taking into account the definitions given above, we estimate the
fraction of stellar mass accreted by galaxies in each magnitude bin at
a given $z$ resulting from the contribution of each of the processes
considered, distinguishing between the two different components, as
\begin{equation}
  {\rm SMFrac}_{b,z} ^{\rm proc, \, comp}= \frac{SM_{b,z} ^{\rm proc,
      \, comp}}{{SM_{b,z} ^{\rm{(i)+(ii)+(iii)}}}}.
\end{equation}
These stellar mass fractions can be grouped according to the aspect we
want to analyse. 

We first consider the fractions of the stellar mass given by new
formed stars, on one hand, and accreted stars, on the other,
regardless of the processes that contribute to them. Thus we have
\begin{equation}  
  {\rm SMFrac}_{b,z} ^{\rm comp}=\sum_{\rm proc=1}^{N_{\rm proc}}{{\rm
      SMFrac}_{b,z} ^{\rm proc, \, comp}},
\end{equation}
for the `stars' and `sat' components.  The dependence of these
fractions with the magnitude bins, in which galaxies along the RS lie,
is presented in Fig. \ref{fig3} for the two components, considering
the values of stellar mass reached at $z=0$.  As can be seen, the
stellar mass received by all galaxies along the RS is mainly composed
by the `stars' contribution ranging from $\sim 95\%$ for the least
luminous galaxies ($M_{T_{1}} \geq -17$, lying in bin $7$), to $\sim
65\%$ for the most luminous ones ($M_{T_{1}} \leq -24$, bin $1$).
This gradual reduction of the fraction of stellar mass contributed by
the stars for the more luminous galaxies occurs at expenses of the
accretion of the stellar component of merging satellites (`sat'
contribution), indicating that mergers become important for the more
massive galaxies.

We now analyse the evolution with redshift of the accumulated stellar
mass contributions from the different processes, within different
magnitudes bins.  The evolution of the corresponding mean fractions of
stellar mass are shown in Fig. \ref{fig4} for galaxies with magnitudes
corresponding to bins $1$, $3$ and $6$; these fractions are given with
respect to the total contribution to the stellar mass of galaxies at
the redshifts considered.  Although we have taken into account both
the `stars' and `sat' components arising from major dry mergers, we
found that the contribution of the former to the whole stellar mass of
galaxies is negligible.  Hence, in the following, we simply refer to
this process as `major dry merger', meaning that only the stellar mass
from the `sat' component is being considered.

From a first inspection of these plots,  we see that quiescent SF
appears as the dominant process that contributes to the stellar mass
of galaxies within all magnitude bins and at all redshifts.  However,
we can see that the corresponding stellar mass fractions decrease
monotonically with redshift as the cold gas reservoir in each galaxy
is exhausted.

The relative contribution of the rest of the processes, broadly
grouped in disc instabilities and mergers, change considerably as we
move along the RS.  The left panel of Fig. \ref{fig4} shows the mean
stellar mass fractions for galaxies with low luminosity ($ -18 <
M_{T_{ 1}} \leq -17$, bin $6$).  We can clearly see that the second
larger contribution to the stellar mass of these galaxies comes from
the stars formed during minor wet mergers (i.e. the `stars'
component); this mass fraction increases with redshift reaching a
value of $\approx 25$ per cent at $z\approx 0$ that has already been
achieved at $z\approx 2$, indicating that minor wet mergers do not
play a significant role since that epoch.  These fractions are
followed by those corresponding to the contribution of the major wet
merger. As we will see, the contribution of this particular process is
very small or almost negligible for more luminous galaxies, therefore,
for the sake of clarity, we do not split this process in the two
different components. The mass contributions provided by disc
instabilities and other kind of mergers are negligible.

As we move up along the RS to higher luminosities, the situation
changes appreciably.  The mass fractions for galaxies in the critical
magnitude bin $3$, where the detachment of the bright end of the RS
from the general linear fit occurs, are shown in the middle panel of
Fig. \ref{fig3}. At these magnitudes, the relevant contribution to the
stellar mass content (apart from quiescent SF) arises from minor dry
mergers, while the relevance of the `stars' component in minor wet
mergers decreases considerably from $z\approx 4$, becoming negligible
at $z=0$, in contrast to the behaviuor manifested by this process for
galaxies in bin $6$.  We also find a mild contribution of major wet
mergers at low redshift. All these merger events are responsible of
$\approx 20\%$ of the stellar mass received by galaxies with
magnitudes in the range $-21 < M_{T_{1}}\leq -20$ at $z=0$.

For the most luminous galaxies ($-24 < M_{T_{1}}\leq -22$, bin $1$),
the trend of the mass contribution of minor dry mergers is reinforced
with respect to the one shown for bin $3$, gaining importance from
earlier epochs ($z\approx 6$), and reaching a higher value of $\approx
22\%$ at $z=0$. The contribution of major dry mergers also increase
notably since $z \approx 2$ giving a fraction of $\approx 10\%$ at
$z=0$. Thus, the whole contribution of these two relevant processes at
the present epoch leads to a fraction of $\approx 32\%$, consistent
with the result shown in Fig.~\ref{fig3}.

\begin{figure}
  \begin{center}
    \includegraphics[width=0.45\textwidth]{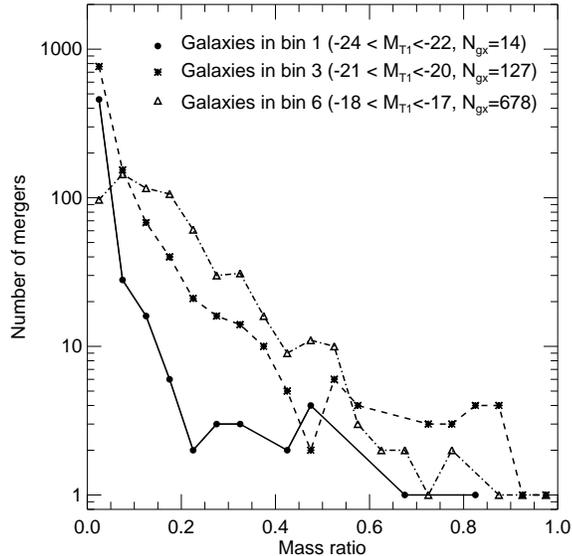}
    \caption{
      Number of mergers suffered by the galaxies in a $\rm{C}15$
      cluster through their whole lifetime as a function of the
      baryonic mass ratio of mergers; different curves correspond to
      magnitude bins $1$, $3$ and $6$, which contain different number
      of galaxies.}
    \label{fig3b}
  \end{center}
\end{figure}

Although the previous description is based on the mean mass fractions
of galaxies in the massive C15 clusters, which clearly show the change
in slope of the RS, similar results are obtained for the small
simulated clusters C14 (not shown). In the latter, however, statistics
become very poor at high luminosities, with only a few objects, if
any, in the highest magnitude bin. This general behaviour is in
agreement with the scenario infered by \citet{Martinez10} from the
analysis of galaxies in groups and in the core of X-ray clusters, in
which these galaxies populate nearly the same CMR. Their result
supports previuos works that indicate that the CMR is independent of
the environmental overdensity  \citep{Hogg04, LopezCruz04}.

\subsection{Mass ratio of mergers resolved in the model}
The contribution of minor mergers to the stellar mass is relevant for
galaxies in all magnitude bins, as it is evident from
Fig.~\ref{fig4}. The mass ratio of minor mergers that are resolved by
the model depends on the mass of the galaxies involved. Given the mass
resolution for DM haloes in the {\em N}-body/SPH simulations used, a
very conservative upper limit for the mass resolution in galaxies
would be $\sim$10$^9 h^{-1}\, M_\odot$ (the minimum resolved halo mass
times the baryonic fraction), but due to the complex interplay of
processes such as gas cooling, feedback and strangulation, we find
that our model produces objects with stellar mass as small as
1.5~$\times$~10$^{5}\, h^{-1} M_\odot$. Therefore, the minimum mass
ratio resolved in the model (corresponding to the merger of such an
object with a BCG, with stellar mass of $\simeq$~1.5~$\times$~10$^{12}
h^{-1}\, M_\odot$ in our simulations) is $\sim 10^{-7}$. However, it
is likely that our model is incomplete for such small objects. 

Comparing the results of the model with e.g. the observed luminosity
function for galaxy clusters \citep{DePropris03}, it seems our model
is complete for galaxies with absolute magnitude $< -15$ in the
$b_J$-band. This corresponds to galaxies with a mean stellar mass
$\simeq$~2~$\times$~10$^8 h^{-1}\, M_\odot$.  Consequently, for the
case of the BCGs, we expect our model to be complete in mergers with
mass ratio $\gtrsim$10$^{-4}$.  For galaxies in the C15 clusters, the
minimum mass ratios resolved in the model are of the order of
4~$\times$~10$^{-5}$, 10$^{-4}$ and 6~$\times$~10$^{-3}$ for bins 1, 3
and 6, respectively. The ratios indeed increase as one considers less
luminous galaxies. When selecting the most luminous galaxies in bin
$1$ in the C14 clusters, the minimum mass ratio is higher but of the
same order of magnitude, $\sim 2.5~\times$~10$^{-5}$.  

In general, the number of mergers suffered by galaxies along their
whole lifetime increases rather monotonically as mergers of lower mass
ratio are considered, as can be seen from Fig. \ref{fig3b}, where the
results are shown for magnitude bins $1$, $3$ and $6$. Minor mergers
(defined as those with mass ratio lower than 0.3; Section
\ref{sec:MSB}) are dominant for all galaxies along the RS. For the
least luminous bin, the maximum is not reached for the least value of
the mass ratio, as in the more luminous ones, reflecting the influence
of the limit in the mass resolution of the simulations.  Each of these
bins contains different number of galaxies, as indicated in the plot,
being larger for the least luminous one. Taking into account this
aspect, the average number of mergers is larger for brighter galaxies.

\subsection{Influence of gas fraction thresholds}
From the analysis of the stellar mass fractions discussed in
Subsection \ref{MassFrac}, it is clear that the evolution of galaxies
in the bright end of the RS (those brighter than the critical value
($M_{T_{1}} \approx -20$) is driven by the dry contribution of both
minor and major mergers. On the other hand, fainter galaxies are
affected by the wet component of mergers.

These conclusions are in agreement with those obtained by
\citet{Skelton09}, who find that most of the galaxies in the bright
end of the CMR of local early-type galaxies have undergone dry
mergers, in contrast to faint galaxies. Their results are based on a
simplified model that isolates the effects of the dry merging in the
colours of the RS galaxies. The merger histories and gas fractions of
galaxies are extracted from the semi-analytic model of
\citet{Somerville08}. Their model is able to reproduce the change of
slope in the bright end of the CMR as seen in local early-type
galaxies from the SDSS. \citeauthor{Skelton09} affirm that this change
in slope, and the magnitude at which the break occurs, depend strongly
on the assumption of the gas fraction threshold below which mergers
are considered dry. This threshold is allowed to vary between~$10$
and~$30$ per cent.

In our model, on the contrary, the tilt of the RS towards bluer
colours has only a weak dependence on the gas thresholds adopted to
distinguish between wet and dry mergers (defined in Section
\ref{sec:MSB} as $f_{\rm gas,minor}$ and $f_{\rm gas,major}$ for the
minor and major mergers, respectively). The parameter $f_{\rm
  gas,major}$ has no influence on the effect of major mergers, because
such mergers always lead to the formation of a spheroid and the
consumption of the cold gas in a starburst, regardless of the amount
of gas available in the merging galaxies. The threshold $f_{\rm
  gas,major}$ only has the purpose of classifying major mergers as
either dry or wet, and thus only affects the percentages of the
`stars' and `sat' components contributed by these processes to the
stellar mass of galaxies.  We have verified that changing this
threshold within the range $0.2 \leq f_{\rm gas,major}\leq 0.6$
results in no significant changes with respect to the results
presented in Figure \ref{fig4}, obtained for $f_{\rm gas,major}=0.4$.
On the other hand, the value $f_{\rm gas,minor}=0.6$ for the minor
mergers has been tuned together with the rest of the free parameters
of the model in order to reproduce several observed galaxy properties,
as described in Section \ref{sec:MSB}.

\begin{figure}
  \begin{center}
    \includegraphics[width=0.45\textwidth]{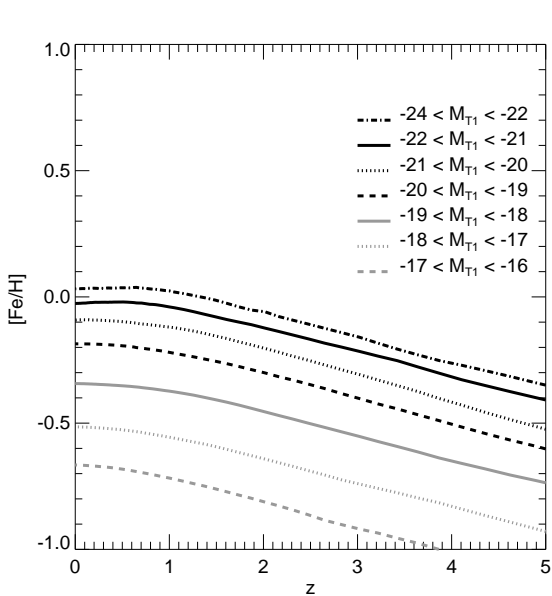}
    \caption{Evolution with redshift of iron abundance of the stellar
      component of galaxies that at $z=0$ lie within a given range of
      magnitudes; metallicity values are a result of the complex
      combination of all processes included in our model SAG. Averages
      at each redshift have been estimated considering galaxies in C15
      clusters. Different lines represent the mean values for galaxies
      within different magnitude bins.}
    \label{fig5}
  \end{center}
\end{figure}

We would like to emphasize that our model was not constructed to
produce a break in the RS; this arises naturally as a consequence of
the physical processes included in the model.  In particular, varying
the value $f_{\rm gas,major}$ over the aforementioned range, a break
in the bright end of our simulated RS is always present, without any
appreciable change in the values of the slope or of the critical break
magnitude.  In this respect, our findings differ from those of
\citet{Skelton09}, who claim to see a strong dependence of the break
on the adopted gas threshold; moreover, we do not see a significant
dependence from their results (see their Fig. $1$).  In any case, both
our results and those of \citet{Skelton09} strongly support the fact
that the dry contribution of the merger processes are relevant in the
formation of the brightest galaxies in the RS, both in field and
cluster environments. It has been suggested that dry mergers are
expected to only increase the mass of the remnant without changing
their colours because there is no associated SF, thus causing the
detachment of the bright end from the general linear trend
(\citealt{Bernardi07, Skelton09}). In what follows, we explain this
claim by means of our detailed model of galaxy formation, focusing on
cluster galaxies.

\begin{figure}
  \begin{center}
    \includegraphics[width=0.45\textwidth]{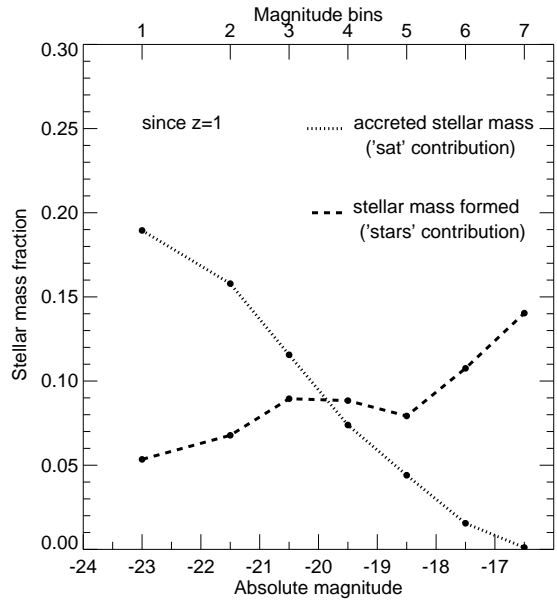}
    \includegraphics[width=0.45\textwidth]{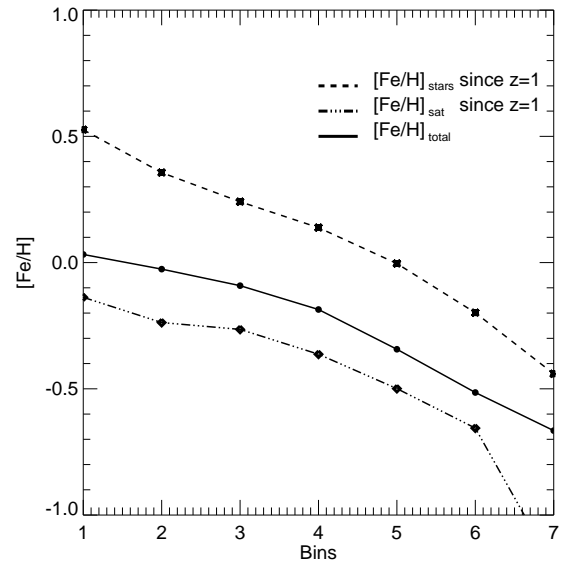}
    \caption{Top panel: stellar mass fractions of `stars' and `sat'
      contributions accumulated since $z=1$ for each magnitude bin,
      normalized with the total stellar mass of galaxies at $z=0$
      within a given bin; galaxies in the C15 clusters are
      considered. Bottom panel: Average iron abundance of galaxies at
      $z=0$ as given by the complete evolution traced by the
      semi-analytic model (solid line), for galaxies within different
      magnitude bins, compared with the metallicity of the mass
      contributed by the stars formed since $z=1$ (dashed line) and of
      the stars already present in the satellites accreted since $z=1$
      (dot-dashed line).}
    \label{fig6}
  \end{center}
\end{figure}

\subsection{Metallicity evolution}
The evolution of the mean value of the stellar iron abundance of
galaxies that at $z=0$ lie within a given range of magnitudes is shown
in Fig.~\ref{fig5}.  These metallicities are the result of the
complete combination of all the processes considered in the
semi-analytic code; averages have been estimated for each magnitude
bin considering the three C15 clusters.  We can see that, regardless
of the galaxy magnitude, the mean metallicity of galaxies increases
smoothly with redshift, being always higher for the more luminous
galaxies. In particular, at $z=0$, galaxies in the bright end
($M_{T_{1}} > -20$) reach metallicity values within a narrower range
($\approx 0.15$ dex) than the rest of the galaxy population. This fact
is directly linked with the similar colours that characterize these
galaxies making them depart form the general trend of the RS. These
chemical abundances seem to be in place since $z=1$.  This is
supported by observations of high redshift clusters, which show that
the slope and scatter in the CMR for morphologically selected ETGs
show little or no evidence of evolution out to $z \approx 1.2$
\citep{Blakeslee03, Mei06, Jaffe10}.

\subsection{Mass assembly and merger rates since $z=1$}
Taking into account that metallicity values of galaxies in the bright
end have already been achieved at $z=1$ (Fig.~\ref{fig5}), we consider
the accumulated mass contributed from both components (`stars' and
`sat') in the redshift range $0<z<1$.  The relative contribution of
these two components are shown in the top panel of Fig.~\ref{fig6} for
galaxies in different magnitude bins.  These mass fractions are
normalized with the total stellar mass of galaxies at $z=0$ within a
given magnitude bin, because we are interested on the relative
importance of the mass accumulated since $z=1$ to the stellar mass of
the galaxy at the present epoch.  Note that this stellar mass is a
result of the whole evolution traced by the semi-analytic code,
including its reduction as a result of the mass recycling arising from
stellar evolution.  However, by construction, the variables used to
analyse the mass fractions shown in Fig. \ref{fig6} (as well as in
Figs. \ref{fig3} and \ref{fig4}) save the information of the mass of
stars borned in a galaxy or added to it during merger events, but they
are not updated to take into account the recycled mass. This fact
specially affects the stellar mass fractions contributed by the mass
of new born stars (`stars' component) since $z=1$, which are shown by
the dashed line in the top panel of Fig. \ref{fig6}.  As we can see,
these fraction ranges from $\approx 0.05$ for the brightest galaxies
to $\approx 0.15$ for the faintest ones.  However, they are in fact
even lower taking into account that they are normalized with the total
stellar mass of galaxies at $z=0$, which have lost $\approx 35$~per
cent of their mass as a result of the evolution of stars belonging to
the high mass tail of a Salpeter IMF (which is the one adopted in our
semi-analytic model).

These fractions of the stellar mass formed in situ since $z=1$, in
which all physical procesesses contributing with cold gas from which
new born stars originate are taken into account, can be compared with
estimations done by \citet{Kaviraj08}. They compute the recent star
formation (RSF; mass fraction of stars that form in the galaxy within
the last 1 Gyr of look-back time in its rest frame) in the high
redshift ($0.5 < z < 0.1$) early-type population from rest-frame UV
and optical colours and combine their results with similar studies in
the local universe \citep{Kaviraj07}, thus being able to estimate the
star formation history of the early-type population between $z=0$ and
$1$.  For the range $0.5<z<1$, early type galaxies belonging to the RS
($NUV-r > 4$) typically show RSF values less than $5$ per cent and the
reddest early-types (which are also the most luminous) are virtually
quiescent with RSF values of $\sim 1$~per cent. 

On the other hand, \citet{Kaviraj07} find that the SF activity in the
most active ETGs (those in the blue peak of the colour-magnitude
distributions) has halved from an average RSF of $\sim 11$ per cent at
$z=0.7$ to $\sim 6$~per cent at the present day.  From these results,
they infer that $10-15$ per cent of the stellar mass in luminous
early-type galaxies ($-23 < M_V < -20.5$) may form after $z=1$, while
less luminous galaxies may form $30-60$ per cent of their mass within
this time-scale.  The fractions of the stellar mass formed in situ
since $z=1$ given by our model (dashed line in Fig.  \ref{fig6}) are a
factor of 2 smaller than these values, within ranges of $\sim 0.05 -
0.09$ and $\sim 0.09 - 0.15$ per cent for galaxies in the bright and
faint end of the RS, respectively.  This difference is expected,
taking into account that we are analysing cluster galaxies where the
peak of the star formation takes place at earlier times, while those
in the study of \citet{Kaviraj08} belong to a variety of
environments. Thus, our results are in very good agreement with these
observational estimates.

The stellar mass fractions contributed by the accreted stellar mass
during mergers are not so affected by the definition of the
accumulated quantities, since the stellar population of satellites
have already suffered from the recycling processes by the time they
are accreted, assuming that the bulk of their stars where formed at $z
> 1$.  Indeed, if a satellite galaxy accreted within the redshift
range $0<z_{\rm merger}<1$ is created as early as $z=7$, then, by the
time it merges with the central galaxy it has already lost $\approx
36$ per cent of its original mass retaining only the low mass (slowly
evolving) stars.  Then, the stars that will be added to the central
galaxy will evolve very slowly and more than 93\% of those stars will
still be alive at $z=0$.  Even in the case of a relatively young
galaxy created at $z=2$ and accreted during the time interval
$0<z_{\rm merger}<1$, we find that more than 90\% of the stellar mass
added during the merger is still alive at $z=0$.  Keeping these
caveats in mind, we can conclude from Fig. \ref{fig6} (upper panel)
that $\approx 10 - 20$ per cent of the mass of galaxies in the bright
end of the RS arises from satellites accreted since $z=1$, with very
few stars being formed in situ.  

Observational quantification of the impact of mergers in the mass
assembly of early type galaxies yields to larger increments of the
mass of galaxies since $z=1$ with respect to what we have
found. However, we have to keep in mind that our results correspond to
cluster galaxies, characterized by a faster evolution, as already
mentioned.  In particular, \citet{Lopez10} find that the brightest
early-type galaxies increase their mass by about $42 \pm 8$ per cent
as a result of mergers since $z=1$.  They use kinematically confirmed
close pairs from the VIMOS VLT deep spectroscopic redshift survey to
study the role of minor mergers to mass assembly of normal $L_B
\gtrsim L_B^{\ast}$ galaxies, which can belong to different
environments. The fact that the mass fraction received by massive
early-type galaxies in their sample is a factor of two larger than the
one estimated for our brightest galaxies is a consequence of the lower
number of mergers that cluster galaxies suffer since $z=1$.

The estimation of the average number of mergers per red galaxies since
$z = 1$ done by \citet{Lopez10} is $1.3\pm 0.3$, that is splitted in
$0.7\pm 0.1$ major mergers and $0.6\pm 0.2$ minor ones.  Their
criterion to separate between minor and major mergers is based on the
ratio of {\em B}-band luminosity of the merging components which is
used as a proxy of mass; minor mergers are characterized by the
luminosity ratio ranging from $0.1$ to $0.25$.  In our model, minor
mergers are defined as those having mass ratios less than $0.3$,
whithout impossing a lower limit. Thus, we find that galaxies in the
bright end of the RS ($M_{T_1}\leq -20$) have suffered an average
number of mergers of $\sim 4.88$ since $z=1$, which is almost four
times larger that the value obtained by \citet{Lopez10}.  Therefore,
we repeat our estimation restricting to minor mergers with mass ratios
larger that $0.1$.  Thus, we find that the average number of mergers
suffered by galaxies in the bright end of the RS is considerably
reduced, with $\sim 0.33$ minor mergers per galaxy.  The value is a
factor of two smaller that the one inferred by \citet{Lopez10}, but
again, we have to keep in mind that our sample contains cluster
galaxies.

The cumulative fraction of stellar mass contributed by minor mergers
since $z=1$ that we compute, ranges from $\sim 0.12$ to $0.07$, for
the most and the least luminous galaxies in the bright end of the RS,
respectively, which are in very good agreement with the estimation of
0.1 done by \citet{Lopez10}.  These values have been obtained in the
same way as those shown in the upper panel of Fig.~\ref{fig6} but
restricting to the contribution of the `sat' component of minor
mergers, including those with mass ratio less than 0.1.  This
indicates that the mass contribution of minor mergers with such a low
mass ratio, altough being highly numerous, contribute with a very
small percentage to the stellar mass of the central galaxy.  On the
other hand, the average number of major merger suffered by the
brightest galaxies in the RS since $z=1$ is $\sim 0.19$, a factor of
three smaller than the one obtained by \citet{Lopez10}, which explains
that the mass fraction contributed by all kind of mergers to the
brightest galaxies in the RS ($\sim 0.2$) is a factor of two smaller
than observed by \citet{Lopez10} ($\sim 0.42$).  However, this value
of the number of major mergers is in better agreement with other
empirically determined major merger rates, which indicates that major
merger activity in galaxies with masses comparable with the ones in
this paper, becomes infrequent after $z=1$ (e.g.
\citealp{Conselice09, Bundy09, Jogee09}).

We also compare our results with those obtained by
\citet{Nierenberg11}. They study the spatial distribution of faint
satellites of early-type galaxies at intermediate redshifts
($0.1<z<0.8$) selected from the GOODS field. Satellites are detected
from high resolution HST images considering those that are up to $5.5$
magnitudes fainter than the host galaxy and as close as $0''5/2.5$ kpc
to the host. The majority of their host galaxies are massive
elliptical galaxies with typical stellar masses of $\approx 10^{10.5}
M_{\odot}$ that have on average $1.7^{+0.9}_{-0.8}$ satellites.  This
inferred satellite number can easily account for the number of minor
mergers ($0.4 - 0.5$) that $60$ per cent of their host galaxies are
expected to undergo in the time span of the study. This number of
mergers has been obtained by using the fitting form for the mean
merger rate per halo as a function of halo mass, progenitor mass ratio
and redshift derived by \citet{Fakhouri10} from halo merger
statistics. Although a bit lower than these infered values, the result
from our model ($\sim 0.33$ minor mergers per galaxies since $z=1$) is
consistent with them.

\subsection{Metal contribution since $z=1$}
An interesting aspect of the plot in the lower panel of
Fig.~\ref{fig6} is that the contributions of the `stars' and `sat'
components cross each other at the break magnitude $M_{T_{1}} = -20$,
clearly indicating that the stellar population accreted during late
mergers determines the properties of galaxies in the bright end of the
RS.  From the discussion related to this plot, we can safely assume
that more than 90\% of the stellar mass formed and added by mergers
since $z=1$ is still alive at $z=0$, determining the metallicity and
colour of galaxies in the RS at the present epoch. We estimate the
metallicity of the stellar mass contributed since $z=1$ by the these
two components, $[\rm{Fe/H}]_{\rm stars}$ and $[\rm{Fe/H}]_{\rm sat}$,
considering the mass of iron, ${\rm Fe}_{g,z}^{\rm proc \, comp}$, and
hydrogen, ${\rm H}_{g,z}^{\rm proc \, comp}$, defined in Section
\ref{Sec:Proc}.  The average value of these metallicities for C15
cluster galaxies within different magnitude bins are shown in the
lower panel of Fig. \ref{fig6}.  They are compared with the average
metallicity achieved by the stellar mass of galaxies at $z=0$
according to the whole evolution traced by the semi-analytic code
(solid line), which lies in the range $-0.65 \lesssim [\rm{Fe/H}]_{\rm
  total} \lesssim 0.05$ for the least to the most luminous galaxies in
the RS.  The metallicity of the new stars formed since $z=1$ (dashed
line) ranges from $[\rm{Fe/H}]_{\rm star}\approx -0.45$ for the least
luminous galaxies to $\approx 0.5$. In particular, for galaxies in the
bright end of the RS, with magnitudes $-24 \leq M_{\rm{T_{1}}}\leq
-20$ (bins $1, 2$ and $3$), the metallicity of the `stars' component
is above solar ($0.25 \lesssim [\rm{Fe/H}]_{\rm stars}\lesssim 0.5$)
as the result of the highly chemically enriched cold gas available for
star formation in more massive galaxies. However, as we have seen, the
mass fraction contributed by processes involving star formation is
very low for these galaxies (see top panel of Fig. \ref{fig6}),
therefore, these metal contributions hardly affect the metallicity of
the whole galaxy. On the other hand, the `sat' component of satellites
accreted since $z=1$ represents an appreciable fraction of the galaxy
mass at $z=0$ for galaxies in the bright end ($\approx 10 - 20$ per
cent), but the values $[\rm{Fe/H}]_{\rm sat}$ (dashed-dotted line in
bottom panel of Fig. \ref{fig6}) are subsolar for galaxies in all
magnitude bins, being $\approx 0.15$ dex lower than the corresponding
galaxy metallicity.  Hence, dry mergers help increase the mass of
galaxies in the bright end of the RS without considerably affecting
their average metallicities. This upper limit in the metallicity
achieved by galaxies in the bright end fixes their colour, being bluer
than expected if star formation from highly chemically enriched gas
were relevant in their final evolution, thus explaining the departure
of the bright end of the RS from a linear fit.

\subsection{Discussion}
Our findings related to the low metallicity of the stellar mass of
satellites accreted al low redshifts ($z<1$), with respect to the
chemical abundances found in the remnant galaxy, are in line with the
results of \citet{Lagos09}. They show a clear correlation between the
metallicity of galaxies and the mass of their host DM haloes, with
central galaxies being more chemically enriched than their surviving
satellites. This trend is a direct consequence of the hierarchical
clustering process, where more massive galaxies sit in initially
higher density peaks and, as a consequence, start to form earlier on
average, allowing them to acquire higher amounts of metals.  

Our results also support the scenario proposed by \citet{Martinez10}
for the formation of the CMR in massive systems. Their analysis of
observational results indicate that the CMR is populated by red ETG
that formed the bulk of their stars during the early stages of massive
halo assembly, on one hand, and red galaxies that passed most of their
lives inhabiting poor groups or the field and fell into massive
systems at lower redshifts, on the other. Our study suggests that part
of the later population are the satellites intervening in the dry
mergers processes, that play a significant role in the late evolution
of massive galaxies.

The fact that massive cluster galaxies are fed by minor dry mergers
whereas fainter galaxies along the CMR are mainly affected by minor
wet ones is due to the influence of the environment in  which the host
galaxy and their satellites reside. More massive galaxies tend to
inhabit denser regions, where it is highly likely that infalling
satellites are stripped of their cold gas before the merger as a
result of environmental effects. Hence, the modelling of the
environmental processes that affect the cold gas content of galaxies
could influence our conclusions regarding the fraction of mass
received by galaxies from dry or wet mergers.  In our model, as in
most previous semi-analytic modelling, the hot gas haloes of galaxies
are completely and immediately stripped as soon as a galaxy becomes a
satellite of a group or cluster. The affected galaxy cannot replenish
its cold gas any further, and with a typical SFR it exhausts its cold
gas in a few Gyr \citep*{Gallagher89}. This process is called
`strangulation' \citep{Larson80}, and provides an explanation for the
SFR gradients in clusters \citep{Balogh00} and the colour evolution of
cluster galaxies \citep{Kodama01}. However, recent work
(e.g. \citealp{Weinmann06}) has shown that, compared to SDSS
observations, semi-analytic models produce higher fractions of red
satellites.  This is likely due to the crude modelling of the removal
of the hot gas haloes. If the hot gas is removed more gradually,
satellite galaxies can prolong their star formation resulting in bluer
colours. It is not known which physical process is actually
responsible for the removal of the hot gas, but one likely candidate
is ram-pressure stripping (RPS), as shown by \citet{McCarthy08} and
\citet{Bekki09}.

This satellite overquenching problem raises the issue that perhaps the
semi-analytic model used overpredicts the number of dry mergers, as
satellite galaxies in the model would be too gas-poor. \citet{Font08}
has shown that considering a gradual removal of hot gas by RPS, a
better match to the observed fractions of red galaxies is
obtained. One would conclude that with such a modification, the number
of wet mergers should increase; however, the situation is not so
simple. If the gas content of satellite galaxies is regulated by RPS,
the colour of the satellite will be bluer as a result of a more
prolonged star formation, but this does not necessarily mean that a
higher mass of gas is contributed to the mergers. In the central
regions of galaxy clusters the ram-pressure exerted by the ICM is
strong enough to significantly affect the gas content of galaxies
\citep{Tecce10} and by the time the satellite merges with the central,
not only its hot gas but also the cold gas could be stripped away by
RPS, becoming part of the ICM. Thus, the gas content of the satellite
would be mostly removed prior to the merger (\citealp{Sofue94};
\citealp*{Grebel03}) and the number of wet mergers might not increase
significantly. We have updated the model described in \citet{Tecce10}
to consider the RPS of the hot gas, and we intend to explore this
issue further in a forthcoming paper.

Finally, it is interesting to note that the break magnitude in which
there is a change of the slope in the RS ($M_{\rm{T_1}} \sim -20$),
corresponds to a galaxy mass of $M_\star \sim 10^{10}\,h^{-1}\,{\rm
  M}_{\sun}$, also marks the difference between two regimes in the
dependence of metallicity gradients with galaxy mass, as found by
\citet{Spolaor09}.  Their results suggest that cluster galaxies above
the mentioned mass threshold might have formed initially by mergers of
gas rich disc galaxies and then subsequently evolved via dry merger
events, in concordance with our findings.

\section {Summary and Conclusions}
\label{Sec:Conclu}
In this work we have studied the physical processes responsible for
the detachment of the bright end of the RS of cluster galaxies from
the linear fit determined by less luminous galaxies. The RS is
constructed by selecting galaxies according to the empirical
redshift-dependent criterion given by \citet{Bell04} to separate the
blue and red sequences in the colour-magnitude diagram. We use the
semi-analytic model of galaxy formation and evolution
\sag~\citep{Lagos08, Tecce10}, combined with cosmological
non-radiative {\em N}-Body/SPH simulations of galaxy clusters in a
concordance $\Lambda$ Cold Dark Matter universe. We consider two sets
of simulated clusters with virial masses in the ranges $\simeq
(1.1-1.2)\times 10^{14}\,h^{-1}\,{\rm M}_\odot$ (C14 clusters) and
$\simeq (1.3-2.3)\times 10^{15}\,h^{-1}\,{\rm M}_\odot$ (C15
clusters).

We find a very good agreement between the general trends of the
simulated RS and observed CMR of ETGs in four different
colour-magnitude planes.  The simulated relation presents a break at
approximately the same magnitude in the different systems ($M_R^{\rm
  break} \sim M_V^{\rm break} \sim M_{T_1}^{\rm break} \approx -20$,
galaxy mass of $\sim 10^{10} {\rm M}_{\odot}$), with more luminous
galaxies having almost constant colours (Washington $C-T_{\rm
  1}\approx 1.5$) (see Fig. \ref{fig1}) instead of following the
linear trend infered from the less luminous ones.

This aspect is consistent with a non-linear fit found from different
set of observations \citep{Baldry06, Janz09, Skelton09}.  The change
of slope in the bright end of the RS emerges naturally from our model
once it has been calibrated to satisfy several observational
constraints simultaneously. One of the obervational relationships that
is closely followed by the simulated galaxies is the
luminosity-metallicity relation (see Fig.~\ref{fig2}).  This fact
supports the use of the chemical history of galaxies as a tool to help
understand the development of the RS and its break at the bright end.

Thus, we focus on the analysis of the mass and metals contributed to
each galaxy by different physical processes: quiescent star formation,
and starbursts during disc instability events and mergers.  We study
the mean properties of galaxies in seven different magnitude bins;
galaxies are grouped according to their $M_{T_1}$ magnitudes.  From
the analysis of the accumulated stellar mass contributed by different
processes, we obtain information about the way in which their relative
importance change with redshift, for galaxies in a given magnitude
bin, and the dependence of their relevance with galaxy luminosity. We
find that:

\begin{itemize}
\item Quiescent SF appears as the dominant process that contributes to
  the stellar mass of galaxies within all magnitude bins and at all
  redshifts. The corresponding stellar mass fractions decrease
  monotonically with redshift as the cold gas reservoir in each galaxy
  is exhausted.
\item Galaxies with low luminosity ($-18 < M_{T_1} \leq -17$, bin 6)
  receive a large contribution from the stars formed during minor and
  major wet mergers, with the former processes giving place to a
  larger mass fraction, which increases with redshift and reaches a
  value of $\approx 25$ per cent at $z\approx 0$. This value has
  already been achieved at $z \approx 2$, indicating that minor wet
  mergers do not play a significant role since that epoch. 
\item The most luminous galaxies ($-24 < M_{T_1} \leq -22$, bin 1)
  have a considerable contribution from minor dry mergers, which
  become important since $z \approx 6$; the corresponding mass
  fraction reaches a value of $\approx 22$ per cent at $z = 0$.  The
  contribution of major dry mergers is also relevant for these
  galaxies since $z \approx 2$, giving a fraction of $\approx 10$~per
  cent at $z = 0$.
\item The transition between the relative importance of the
  contribution of wet and dry mergers occurs for galaxies within the
  range of magnitudes ($-21 < M_{T_1} \leq -20$, bin 3; $M_\star \sim
  10^{10} {\rm M}_{\odot}$), that is, galaxies characterized by the
  break magnitude, where the detachment towards bluer colours in the
  RS takes place.  At these magnitudes, minor dry mergers start to
  gain importance, while the relevance of the stars formed during
  starbursts in minor wet mergers decreases considerably from $z
  \approx 4$, becoming negligible at $z = 0$, in contrast to the
  behaviuor manifested by this process for the least luminous galaxies
  (bin 6). We also find a mild contribution of major wet mergers at
  low redshifts.  All these merger events are responsible of $\approx
  20$ per cent of the total mass received by galaxies.
\end{itemize}

From the analysis of the relative importance of the stellar mass
fractions contributed by different processes, it is clear that the
evolution of galaxies at the bright end of the RS is driven by the dry
contribution of both minor and major mergers, while fainter galaxies
are affected by the wet component of mergers.  This situation is
already established at $z\approx 1$. The metallicities of galaxies in
the bright end do not change since that redshift, although a mild
increase is detected for lower luminous galaxies where star formation
from chemically enriched gas keeps taking place.  

Considering the mass accumulated since $z=1$ from both the stars
formed in situ and and the stellar mass already present in the
accreted satelites, we find that the latter represents a fraction of
the present stellar mass of galaxies in the bright end of the RS that
ranges between $\approx 10$~and $20$~per cent. This fraction is larger
than the fraction of stars formed during recent star
formation. Altough the metallicity of the stars formed since $z\approx
1$ is high ($\approx 0.2 - 0.5$ dex above the mean iron abundances of
galaxies at $z=0$, see Fig. \ref{fig6}, bottom panel), their mass
contribution is so low ($\approx 0.05$) that they do not affect the
mean metallicity of galaxies.  On the contrary, dry mergers provide a
higher fraction of stellar mass than that generated by recent SF but
with low content of metals ($\approx 0.2$~dex below the mean
metallicity of galaxies) thus, their contribution only increase the
stellar mass of galaxies without changing their mean
metallicities. Hence, galaxies in the bright end, which have
negligible star formation activity since $z\approx 1$, reach an upper
limit in their abundances that in turn fixes their colours, taking
into account the small spread in age of these systems.  The departure
of the bright end of the RS from a linear fit can be explained by the
effect of the contribution of minor and major dry mergers, which
increase the mass/luminosity of the galaxies but without changing
their colours.

These conclusions are valid for both high and low mass galaxy
clusters. We have demonstrated that the local properties of galaxies
on the RS seem to be established since $z\approx 1$. A detailed
analysis of the features of the RS, such as its dependence with
environment and redshift, will be addressed in a series of forthcoming
papers.

\section{Acknowledgements}
\vskip -.1cm
We are very grateful to the anonymous referee for a comprehensive and
insightful report which helped to impove this paper. We warmly thank
Marcelo M. Miller Bertolami and Nelson D. Padilla for largely
contributing with useful suggestions and comments. We thank Klaus
Dolag for making the simulations available to us, and Scott Trager for
kindly providing us his observational data on ETGs. We appreciate the
technical support offered by Ruben E. Mart\'inez, Pablo Santa Mar\'ia
and Cristian Vega.  This work was supported by Consejo Nacional de
Investigaciones Cient\'ificas y T\'ecnicas, Agencia de Promoci\'on
Cient\'ifica y Tecnol\'ogica, National University of La Plata,
Argentina and Institute of Astrophysics La Plata (IALP).

\label{lastpage}


\begin{thebibliography}{}

 
\bibitem[\protect\citeauthoryear{Adelman-McCarthy et
    al.}{2008}]{Adelman08}Adelman-McCarthy J., et al., 2008, ApJS,
  175, 297

\bibitem[\protect\citeauthoryear{Asplund, Grevesse \& Sauval}{Asplund
    et al.}{2005}]{Asplund05}Asplund M., Grevesse N., Sauval A. J.,
  2005, in T.G. Barnes III, F.N. Bash, eds, ASP Conf. Ser. Vol. 36,
  Cosmic Abundances as Records of Stellar Evolution and
  Nucleosynthesis. Astron. Soc. Pac., San Francisco, p. 25
 
\bibitem[\protect\citeauthoryear{Baldry et al.}{2004}]{Baldry04}Baldry
  I. K., Glazebrook K., Brinkmann J., Ivezi\'c Z., Lupton R. H.,
  Nichol R. C., Szalay A. S., 2004, ApJ, 600, 681

\bibitem[\protect\citeauthoryear{Baldry et al.}{2006}]{Baldry06}Baldry
  I. K., Balogh M. L., Bower R. G., Glazebrook K., Nichol R. C.,
  Bamford S. P., Budavari T., 2006, MNRAS, 373, 469

\bibitem[\protect\citeauthoryear{Balogh, Navarro \& Morris}{Balogh et
    al.}{2000}]{Balogh00}Balogh M. L., Navarro J. F., Morris S. L.,
  2000, ApJ, 540, 113

\bibitem[\protect\citeauthoryear{Bezanson et
    al.}{2009}]{BezansonvanDokkum10} Bezanson R., van Dokkum P. G.,
  Tal T., Marchesini D., Kriek M., Franx M., Coppi P., 2009, Apj, 697,
  1290

\bibitem[\protect\citeauthoryear{Bournaud, Jog \&
    Combes}{2007}]{Bournaud07}Bournaud F., Jog C. J., Combes F., 2007,
  A\&A, 476, 1179

\bibitem[\protect\citeauthoryear{Bekki}{2009}]{Bekki09} Bekki K.,
  2009, MNRAS, 399, 2221

\bibitem[\protect\citeauthoryear{Bell et al.}{2004}]{Bell04}Bell
  F. E., Wolf C., Meisenhimer K., Rix H.-W. et al., 2004, ApJ, 608,
  752

 
\bibitem[\protect\citeauthoryear{Bell et al.}{2006a}]{Bell06a}Bell
  F. E., Naab T., McIntosh D. H. et al., 2006, ApJ, 640, 241

\bibitem[\protect\citeauthoryear{Bell et al.}{2006b}]{Bell06b}Bell
  F. E., Phleps S., Somerville R., Wolf C., Borch A., Meisenheime K.,
  2006, ApJ, 652, 270
 
\bibitem[\protect\citeauthoryear{Bernardi et
    al.}{2007}]{Bernardi07}Bernardi M., Hyde J. B., Sheth R. K.,
  Miller J. C., Nichol R. C., 2007, ApJ, 133, 1741

\bibitem[\protect\citeauthoryear{Bernardi et al.}{2005}]{Bernardi05}
  Bernardi M., Sheth R. K., Nichol R. C., Schneider D. P., Brinkmann
  J., 2005, AJ, 129, 61

\bibitem[\protect\citeauthoryear{Bernardi}{2009}]{Bernardi09} Bernardi
  M., 2009, MNRAS, 395, 1491

\bibitem[\protect\citeauthoryear{Bernardi et al.}{2010}]{Bernardi10}
  Bernardi M., Roche N., Shankar F., Sheth R. K., 2010, MNRAS,
  submitted (arXiv:1005.3770)
\bibitem[\protect\citeauthoryear{Bessel}{2001}]{Bessel01}Bessell
  M. S., 2001, PASP, 113, 66
 
\bibitem[\protect\citeauthoryear{Bower, Lucey \& Ellis}{Bower et
    al.}{1992}]{Bower92}Bower R., Lucey J., Ellis R., 1992, MNRAS,
  254, 601

\bibitem[\protect\citeauthoryear{Bower, Kodama \& Terlevich}{1998}]{Bower98}
Bower R. G., Kodama T., Terlevich A., 1998, MNRAS, 299, 1193

\bibitem[\protect\citeauthoryear{Blakeslee et al.}{2003}]{Blakeslee03}
  Blakeslee J. P., Franx M., Postman M., Rosati P., Holden B.,
  Illingworth G. D., Ford H. C., Cross N. J. G., et al., 2003, ApJL,
  596, 143

\bibitem[\protect\citeauthoryear{Bruzual \&
    Charlot}{2003}]{Bruzual03}Bruzual G., Charlot S., 2003, MNRAS,
  344, 1000

\bibitem[\protect\citeauthoryear{Bundy et al.}{2009}]{Bundy09}Bundy
  K., Fukugita M., Ellis R. S., Targett T. A., Belli S., Kodama T.,
  2009, ApJ, 697, 1369


\bibitem[\protect\citeauthoryear{Cappellari et
    al.}{2009}]{Cappellari09} Cappellari Michele; di Serego Alighieri
  S., Cimatti A., ApJL, 2009, 704, 34

\bibitem[\protect\citeauthoryear{Cenarro \&
    Trujillo}{2009}]{CenarroTrujillo09} Cenarro A. J., Trujillo I.,
  2009, ApJL, 696, 43

\bibitem[\protect\citeauthoryear{Conselice}{2006}]{Conselice06}Conselice
  C., 2006, MNRAS, 373, 1389

\bibitem[\protect\citeauthoryear{Conselice et
    al.}{2003}]{Conselice03}Conselice C., Gallagher J. S., Wyse R. F.,
  2003, AJ, 125, 66

\bibitem[\protect\citeauthoryear{Conselice et
    al.}{2009}]{Conselice09}Conselice C., Yang C..,Bluck A. F., 2009,
  MNRAS, 394, 1956

\bibitem[\protect\citeauthoryear{Cora}{2006}]{Cora06}Cora S. A., 2006,
  MNRAS, 368, 1540

\bibitem[\protect\citeauthoryear{Croton et al.}{2006}]{Croton06}Croton
  D. J., Springel V., White S. D. M., et al., 2006, MNRAS, 365, 11

\bibitem[\protect\citeauthoryear{de Vaucouleurs}{1961}]{Vacu61}de
  Vaucouleurs G., 1961, ApJS, 5, 233
	
\bibitem[\protect\citeauthoryear{Daddi et al.}{2005}]{Daddi05} Daddi
  E., Renzini A., Pirzkal N., et al., 2005, ApJ, 626, 680

\bibitem[\protect\citeauthoryear{De Lucia \&
    Blaizot}{2007}]{DeLucia07} De Lucia G., Blaizot J., 2007, MNRAS,
  375, 2

\bibitem[\protect\citeauthoryear{DePropris et al.}{2003}]{DePropris03}
  De Propris R., Colless M., Driver S. P., Couch W., Peacock J. A.,
  Baldry I., Baugh C. M., Bland-Hawthorn J., et al., 2003, MNRAS, 342,
  725

\bibitem[\protect\citeauthoryear{Dirsch, Richtler \& Bassino} {Dirsch
    et al.}{2003}]{Dirsch03}Dirsch B., Richtler T., Bassino L.P.,
  2003, A\&A, 408, 929

\bibitem[\protect\citeauthoryear{Dolag et al.}{2005}]{Dolag05}Dolag
  K., Vazza F., Brunetti G., Tormen G., 2005, MNRAS, 364, 753

\bibitem[\protect\citeauthoryear{de Rijcke et
    al.}{2009}]{deRijcke09}de Rijcke S., Penny S. J., Conselice C. J.,
  Valcke S., Held E. V., 2009, MNRAS, 393, 798

\bibitem[\protect\citeauthoryear{Eliche-Moral et
    al.}{2010}]{Moral10}Eliche-Moral M., Prieto M., Gallego J.,
  Zamorano J., 2010, ApJ, submitted (arXiv:1003.0686)

\bibitem[\protect\citeauthoryear{Fan et al.}{2010}]{Fan10} Fan L.,
  Lapi A., Bressan A., Bernardi M., De Zotti G., Danese L., 2010, ApJ,
  718, 1460

\bibitem[\protect\citeauthoryear{Fakhouri et al.}{2010}]{Fakhouri10}
Fakhouri O., Ma C.-P., Boylan-Kolchin M., 2010, MNRAS, 406, 2267

\bibitem[\protect\citeauthoryear{Ferrarese et
    al.}{2006}]{LaFerrarese06}Ferrarese L., Cot\'e P., Jordan A. et
  al., 2006, ApJS, 164, 334

\bibitem[\protect\citeauthoryear{Font et al.}{2008}]{Font08} Font
  A. S. et al., 2008, MNRAS, 389, 1619

\bibitem[\protect\citeauthoryear{Forbes \&
    Forte}{2001}]{ForbesForte01}Forbes D., Forte J. C., 2001, MNRAS,
  322, 257

\bibitem[\protect\citeauthoryear{Forte, Faifer \& Geisler}{Forte et
    al.}{2007}]{Forte2007}Forte J. C., Faifer F. R., Geisler D., 2007,
  MNRAS, 382, 1947

\bibitem[\protect\citeauthoryear{Fukugita, Shimasaku \&
    Ichikawa}{Fukugita et al.}{1995}]{Fukugita95} Fukugita M.,
  Shimasaku K., Ichikawa T., 1995, PASP, 107, 945

\bibitem[\protect\citeauthoryear{Gallagher, Hunter \&
    Bushouse}{Gallagher et al.}{1989}]{Gallagher89} Gallagher J.,
  Hunter D., Bushouse H., 1989, AJ, 97, 700

\bibitem[\protect\citeauthoryear{Gallazzi et al.}{2006}]{Gallazzi06}
Gallazzi A., Charlot S., Brinchmann J., White S. D. M., 2006, MNRAS, 370, 1106

\bibitem[\protect\citeauthoryear{Geisler}{1996}]{Geisler96}Geisler D.,
  1996, AJ, 111, 480

\bibitem[\protect\citeauthoryear{Grebel, Gallagher \& Harbeck}{Grebel
    et al.}{2003}]{Grebel03} Grebel E. K., Gallagher J. S. III,
  Harbeck D., 2003, AJ, 125, 1926

\bibitem[\protect\citeauthoryear{Harris \&
    Harris}{2002}]{Harris2002}Harris W. E., Harris G. L., 2002, AJ,
  123, 3108

\bibitem[\protect\citeauthoryear{Hernquist \&
    Mihos}{1995}]{HM95}Hernquist, L. , Mihos, J. C. , 1995, ApJ , 448,
  41H

\bibitem[\protect\citeauthoryear{Hilker, Mieske \& Infante}{Hilker et
    al.}{2003}]{Hilker03}Hilker M., Mieske S., Infante L., 2003, A\&A,
  397, L9

\bibitem[\protect\citeauthoryear{Hogg et al.}{2004}]{Hogg04} Hogg,
  D. W., et al. 2004, ApJL, 601, L29

\bibitem[\protect\citeauthoryear{Hopkins et
    al.}{2009}]{Hopkins09}Hopkins P. F., Hernquist L., Cox T. J.,
  Keres D., Wuyts S., 2009, ApJ, 691, 1424

\bibitem[\protect\citeauthoryear{Jaff et al.}{2010}]{Jaffe10} Jaff,
  Y. L., Aragon-Salamanca A., De Lucia G., Jablonka P., Rudnick G.,
  Saglia R., Zaritsky D., 2010, MNRAS, in press (arXiv:1007.1425)
 

\bibitem[\protect\citeauthoryear{Janz \& Lisker}{2009}]{Janz09}Janz
  J., Lisker T., 2009, ApJ, 696, L102

\bibitem[\protect\citeauthoryear{Janz \& Lisker}{2008}]{Janz08}Janz
  J., Lisker T., 2008, ApJ, 689, L25


\bibitem[\protect\citeauthoryear{Jogee et al.}{2009}]{Jogee09}
Jogee S., Miller S. H. , Penner K. , et al., 2009, Apj, 697, 1971

\bibitem[\protect\citeauthoryear{Kaviraj et al.}{2005}]{Kaviraj05}
Kaviraj S., Devriendt J. E. G., Ferreras I., Yi S. K., 2005, MNRAS, 360, 60

\bibitem[\protect\citeauthoryear{Kaviraj et al.}{2007}]{Kaviraj07}
  Kaviraj S., Schawinski K., Devriendt J. E. G., Ferreras I., Khochfar S.,
 Yoon S.-J., Yi S. K., Deharveng J.-M., et al., 2007, APJS, 173, 619 

\bibitem[\protect\citeauthoryear{Kaviraj et al.}{2008}]{Kaviraj08}
  Kaviraj S., Khochfar S., Schawinski K., Yi S. K., Gawiser E., 
  et al., 2008, MNRAS, 388, 67

\bibitem[\protect\citeauthoryear{Kaviraj et al.}{2011}]{Kaviraj11}
Kaviraj S., Tan K.-M., Ellis R. S., Silk J., 2011, MNRAS, 411, 2148

\bibitem[\protect\citeauthoryear{Karick, Drinkwater \& Gregg}{Karick
    et al.}{2003}]{Karick03}Karick A. M., Drinkwater M. J., Gregg M.,
  2003, MNRAS, 344, 188

\bibitem[\protect\citeauthoryear{Kauffmann et
    al.}{1999}]{Kauff99}Kauffmann G., Colberg J. M., Diaferio A.,
  White S. D. M., 1999, MNRAS, 303, 188

\bibitem[\protect\citeauthoryear{Kawata \&
    Mulchaey}{2008}]{Kawata08}Kawata D., Mulchaey J. S., 2008, ApJ,
  672, L103

\bibitem[\protect\citeauthoryear{Khochfar \& Burkert}{2003}]{KhochfarBurkert03}
Khochfar  S., Burkert  A., 2003, ApJL, 597, 117 

\bibitem[\protect\citeauthoryear{Khochfar \&
    Burkert}{2006}]{Khochfar06a}Khochfar S., Burkert A., 2006, A\&A,
  445, 403

\bibitem[\protect\citeauthoryear{Khochfar \&
    Silk}{2006}]{Khochfar06b}Khochfar S., Silk J., 2006, ApJ, 648, L21

\bibitem[\protect\citeauthoryear{Kodama \& Bower}{2001}]{Kodama01}
  Kodama T., Bower R., 2001, MNRAS, 321, 18

\bibitem[\protect\citeauthoryear{Lagos, Cora \& Padilla}{Lagos et
    al.}{2008}]{Lagos08} Lagos C., Cora S. A., Padilla N. D., 2008,
  MNRAS, 388, 587

\bibitem[\protect\citeauthoryear{Lagos, Padilla \& Cora}{Lagos et
    al.}{2009}]{Lagos09} Lagos C., Padilla N. D., Cora S. A., 2009,
  MNRAS, 397, L31

\bibitem[\protect\citeauthoryear{Lanzoni et al.}{2005}]{Lanzoni05}
Lanzoni B., Guiderdoni B., Mamon G. A., Devriendt J., Hatton S.,
2005, MNRAS, 361, 369 

\bibitem[\protect\citeauthoryear{Larson, Tinsley, \& Caldwell}{Larson
    et al.}{1980}]{Larson80}Larson R. B., Tinsley B. M., Cadwell
  C. N., 1980, ApJ, 237, 692

\bibitem[\protect\citeauthoryear{Lin et al.}{2010}]{Lin10}Lin L.,
  Cooper M. C., Jian H.-Y. et al., 2010, ApJ, 718, 1158

\bibitem[\protect\citeauthoryear{Lisker, Grebel \& Binggeli}{Lisker et
    al.}{2008}]{Lisker08}Lisker T., Grebel E. K., Binggeli B., 2008,
  AJ, 135, 380

\bibitem[\protect\citeauthoryear{Lopez-Sanjuan et at.}{2010}]{Lopez10}
  Lopez-Sanjuan,C., Le Favre, O., de Ravel, L., Cucciati, O., Ilbert,
  O., Tresse, L., et al. 2010, accepted for publication in A\&A


\bibitem[\protect\citeauthoryear{Liu et al.}{2009}]{Liu09}
Liu F. S., Mao S., Deng Z. G., Xia X. Y., Wen Z. L., 2010, MNRAS, 396, 2003 

\bibitem[\protect\citeauthoryear{L\'opez-Cruz, Barkhouse \&
    Yee}{L\'opez-Cruz et al.}{2004}]{LopezCruz04}L\'opez-Cruz O.,
  Barkhouse W. A., Yee H. K. C., 2004, ApJ, 614, 679

\bibitem[\protect\citeauthoryear{Malbon et al.}{2007}]{Malbon07}
Malbon R. K., Baugh C. M., Frenk C. S., Lacey C. G., 2007, MNRAS, 382, 1394

\bibitem[\protect\citeauthoryear{Mancini et al.}{2010}]{Mancini10}
  Mancini C., Daddi E., Renzini A., Salmi F., McCracken H. J., Cimatti
  A., Onodera M., Salvato M., et al., 2010, MNRAS, 401, 933

\bibitem[\protect\citeauthoryear{Mart\'inez et al.}{2010}]{Martinez10}
  Martinez H. J., Coenda V., Muriel H., MNRAS, 403, 748

\bibitem[\protect\citeauthoryear{Mei et al.}{2009}]{Mei09}
Mei S., Holden B., Blakeslee J. P., Ford H. C., Franx M., Homeier N. L.,
Illingworth G. D., Jee M. J., et al., 2009, APJ, 690, 42 

\bibitem[\protect\citeauthoryear{Menci et al.}{2008}]{Menci08}
Menci N., Rosati P., Gobat R., Strazzullo V., Rettura A., Mei S., Demarco R.,
2008, ApJ, 685, 863 

\bibitem[\protect\citeauthoryear{McCarthy et al.}{2008}]{McCarthy08}
  McCarthy I. G. et al., 2008, MNRAS, 383, 593

\bibitem[\protect\citeauthoryear{McIntosh et
    al.}{2008}]{McIntosh08}McIntosh D. H., Guo Y., Hertzberg J., Katz
  N., Mo H. J., van den Bosch F. C., Yang X., 2008, MNRAS, 388, 1537

\bibitem[\protect\citeauthoryear{Mei et al.}{2006}]{Mei06} Mei S.,
  Holden B. P., Blakeslee J. P., Rosati P., Postman M., Jee M. J.,
  Rettura A., Sirianni M., et al., 2006, ApJ, 644, 759

\bibitem[\protect\citeauthoryear{Mendel et al.}{2009}]{Mendel09}
Mendel J. T., Proctor R. N., Rasmussen J., Brough S., Forbes D. A., 2009,
396, 2103

\bibitem[\protect\citeauthoryear{Mieske et al.}{2007}]{Mieske07}Mieske
  S., Hilker M., Infante L., Mendes de Oliveira C., 2007, A\&A, 463,
  503

\bibitem[\protect\citeauthoryear{Misgeld, Mieske \& Hilker}{Misgeld et
    al.}{2008}]{Misgeld08}Misgeld I., Mieske S., Hilker M., 2008,
  A\&A, 486, 697

\bibitem[\protect\citeauthoryear{Misgeld, Hilker \& Mieske}{Misgeld et
    al.}{2009}]{Misgeld09}Misgeld I., Hilker M., Mieske S., 2009,
  A\&A, 496, 683

\bibitem[\protect\citeauthoryear{Moore et al.}{1999}]{Moore99}Moore
  B., Lake G., Quinn T., Stadel J., 1999, MNRAS, 304, 465
		
\bibitem[\protect\citeauthoryear{Naab et al.}{2007}]{Naab07} Naab T.,
  Johansson P. H., Ostriker J. P., Efstathiou G., 2007, ApJ, 658, 710

\bibitem[\protect\citeauthoryear{Naab, Johansson \& Ostriker}{Naab et
    al.}{2009}]{Naab09}Naab T., Johansson P. H., Ostriker J. P., 2009,
  ApJ, 699, L178

\bibitem[\protect\citeauthoryear{Nakazawa et
    al.}{2000}]{Nakazawa00}Nakazawa K., Makishima K., Fukazawa Y.,
  Tamura T., 2000, PASJ, 52 ,623


\bibitem[\protect\citeauthoryear{Nierenberg et
    al.}{2011}]{Nierenberg11}Nierenberg, A. M.; Auger, M. W.; Treu,
  T.; Marshall, P. J.; Fassnacht, C. D., 2011,ApJ, 731, 44N

\bibitem[\protect\citeauthoryear{Pedersen, Yoshii \&
    Sommer-Larsen}{Pedersen et al.}{1997}]{Pedersen97}Pedersen K.,
  Yoshii Y., Sommer-Larsen J., 1997, ApJ, 485, L17

\bibitem[\protect\citeauthoryear{Reda et al.}{2004}]{Reda04}Reda F.,
  Forbes D., Beasley M., O'Sullivan E., Goudfrooij P., 2004, MNRAS,
  354, 851

\bibitem[\protect\citeauthoryear{Reda et al.}{2005}]{Reda05}Reda F.,
  Forbes D., Hau G., 2005, MNRAS, 360, 693

\bibitem[\protect\citeauthoryear{Robaina et al.}{2010}]{Robaina10}
  Robaina A. R., Bell E. F., van der Wel A., Somerville R. S., Skelton
  R. E., McIntosh D. H., Meisenheimer K., Wolf C., 2010, ApJ, 719, 844

\bibitem[\protect\citeauthoryear{Roche et al.}{2010}]{Roche10}

\bibitem[\protect\citeauthoryear{Roediger}{2009}]{Roediger09}Roediger
  E., 2009, AN, 330, 888

\bibitem[\protect\citeauthoryear{Romeo et al.}{2008}]{Romeo08}
Romeo A. D., Napolitano N. R., Covone G., Sommer-Larsen J.,
Antonuccio-Delogu V., Capaccioli M., 2008, MNRAS, 389, 13

\bibitem[\protect\citeauthoryear{Ruszkowski \&
    Springel}{2009}]{Ruszkowski09}Ruszkowski M., Springel V., 2009,
  ApJ, 696, 1094

\bibitem[\protect\citeauthoryear{Sandage \&
    Visvanathan}{1978}]{Sandage78}Sandage A., Visvanathan N., 1978,
  ApJ, 223, 70

\bibitem[\protect\citeauthoryear{Saro et al.}{2006}]{Saro06}
Saro A., Borgani S., Tornatore L., Dolag K., Murante G., Biviano A.,
Calura F., Charlot S., 2006, MNRAS, 373, 397

\bibitem[\protect\citeauthoryear{Secker, Harris \& Plummer}{Secker et
    al.}{1997}]{Secker97}Secker J., Harris W. E., Plummer J. D., 1997,
  PASP, 109, 1377

\bibitem[\protect\citeauthoryear{Skelton, Bell \& Somerville}{Skelton
    et al.}{2009}]{Skelton09}Skelton R., Bell E., Somerville R., 2009
  , ApJ, 699, L9

\bibitem[\protect\citeauthoryear{Smith Castelli et
    al.}{2008}]{Analia08}Smith Castelli A. V., Bassino L. P., Richtler
  T., Cellone S., Aruta C., Infante L., 2008, MNRAS, 386, 2311

\bibitem[\protect\citeauthoryear{Sofue}{1994}]{Sofue94} Sofue Y.,
  1994, ApJ, 423, 207


\bibitem[\protect\citeauthoryear{Somerville et
    al.}{2008}]{Somerville08}Somerville R.S., Hopkins P. F., Cox
  T. J., Robertson B. E., Hernquist L., 2008, MNRAS, 391, 481

\bibitem[\protect\citeauthoryear{Springel et
    al.}{2001}]{Springel01}Springel V., White S., Tormen G., Kauffmann
  G., 2001, MNRAS, 328, 726

\bibitem[\protect\citeauthoryear{Springel}{2005}]{Springel05b}Springel
  V., 2005, MNRAS, 364, 1105

\bibitem[\protect\citeauthoryear{Springel et
    al.}{2005}]{Springel05a}Springel V., White S. D. M., Jenkins A. et
  al., 2005, Natur, 435, 629

\bibitem[\protect\citeauthoryear{Spolaor et al.}{2009}]{Spolaor09}
  Spolaor M., Proctor R. N., Forbes D. A., Couch W. J., 2009, ApJL,
  691, 138

\bibitem[\protect\citeauthoryear{Stott et al.}{2009}]{Stott09}
Stott J. P., Pimbblet K. A., Edge A. C., Smith G. P., Wardlow J. L., 2009,
MNRAS,394, 2098 

\bibitem[\protect\citeauthoryear{Tecce et al.}{2010}]{Tecce10}Tecce
  T. E., Cora S. A., Tissera P. B., Abadi M. G., Lagos C. del P.,
  2010, MNRAS, 408, 2008

\bibitem[\protect\citeauthoryear{Tormen, Bouchet \& White}{Tormen et
    al.}{1997}]{Tormen97}Tormen G., Bouchet F., White S. D. M., 1997,
  MNRAS, 286, 865

\bibitem[\protect\citeauthoryear{Trager et al.}{2000}]{Trager00}
  Trager S. C., Faber S. M., Worthey G., Gonzalez J. J., 2000a, AJ,
  120, 165

\bibitem[\protect\citeauthoryear{Trager et al.}{2008}]{Trager08}
Trager S. C., Faber S. M., Dressler A., 2008, MNRAS, 386, 715

\bibitem[\protect\citeauthoryear{van der Wel et
    al.}{2009}]{vanderWel09}van der Wel A., Bell E. F., van den Bosch
  F. C., Gallazzi A., Rix H.-W., 2009, ApJ, 698, 1232

\bibitem[\protect\citeauthoryear{Visvanathan \&
    Sandage}{1977}]{Visvanathan77}Visvanathan N., Sandage A., 1977,
  ApJ, 216, 214

\bibitem[\protect\citeauthoryear{Whitaker \& van
    Dokkum}{2008}]{WhitakervanDokkum08} Whitaker K. E., van Dokkum
  P. G., 2008, ApJL, 676, 105

\bibitem[\protect\citeauthoryear{van Dokkum et
    al.}{2010}]{vanDokkum10} van Dokkum P. G., Whitaker K. E., Brammer
  G., Franx M., Kriek M., Labbe I., Marchesini D., Quadri R., et al.,
  2010, Apj, 709, 1018

\bibitem[\protect\citeauthoryear{Weinmann et al.}{2006}]{Weinmann06}
  Weinmann S. M., van den Bosch F. C., Yang X., Mo H. J., 2006, MNRAS,
  366, 2

\bibitem[\protect\citeauthoryear{Yoshida, Sheth \& Diaferio}{Yoshida
    et al.}{2001}]{Yoshida01}Yoshida N., Sheth R. K., Diaferio A.,
  2001, MNRAS, 328, 669


\end{thebibliography}
\end{document}